\documentclass[twocolumn,pra,superscriptaddress]{revtex4-1}

\usepackage[utf8]{inputenc}
\usepackage[english]{babel}
\usepackage[T1]{fontenc}
\usepackage{lmodern}
\usepackage{graphicx}
\usepackage{amsmath}
\usepackage{amsfonts}
\usepackage{amssymb}
\usepackage{microtype}
\usepackage{braket}
\usepackage{subfigure}

\usepackage[breaklinks=true,colorlinks]{hyperref}

\DeclareMathOperator{\sech}{sech}

\newcommand{\lu}{\mathcal{L}_\mathrm{U}}
\newcommand{\lt}{\mathcal{L}_\mathrm{T}}
\newcommand{\w}{\omega}

\newcommand{\la}{\langle}
\newcommand{\ra}{\rangle}
\newcommand{\Tc}{T_\mathrm{c}}
\newcommand{\Th}{T_\mathrm{h}}
\newcommand{\Qc}{Q_\mathrm{c}}
\newcommand{\Qh}{Q_\mathrm{h}}
\newcommand{\ah}{\alpha_\mathrm{h}}
\newcommand{\ac}{\alpha_\mathrm{c}}
\newcommand{\rt}{\rho_\mathrm{tot}}
\newcommand{\Hun}{H_\mathrm{un}}
\newcommand{\Hk}{H_\mathrm{k}}
\newcommand{\omegaun}{\omega_\mathrm{un}}
\newcommand{\omegak}{\omega_\mathrm{k}}
\newcommand{\Tr}{\operatorname{Tr}}

\newcommand{\realp}{\operatorname{Re}}

\begin{document}

\title{Quantum magnetometry using two-stroke thermal machines}

\author{Sourav Bhattacharjee}
\affiliation{Department of Physics, IIT Kanpur, Kanpur-208016, India}
\author{Utso Bhattacharya}
\affiliation{Department of Physics, IIT Kanpur, Kanpur-208016, India}
\affiliation{ICFO - Institut de Ciències Fotòniques, The Barcelona Institute of Science and Technology,
Av. Carl Friedrich Gauss 3, 08860 Castelldefels (Barcelona), Spain}
\author{Wolfgang Niedenzu}
\affiliation{Institut f\"ur Theoretische Physik, Universit\"at Innsbruck, Technikerstra{\ss}e~21a, A-6020~Innsbruck, Austria}
\author{Victor Mukherjee}
\affiliation{Department of Physical Sciences, IISER Berhampur, Berhampur 760010, India}
\author{Amit Dutta}
\email{dutta@iitk.ac.in}
\affiliation{Department of Physics, IIT Kanpur, Kanpur-208016, India}

\date{\today}

\begin{abstract}
  The precise estimation of small parameters is a challenging problem in quantum metrology. Here, we introduce a protocol for accurately measuring weak magnetic fields using a two-level magnetometer, which is coupled to two (hot and cold) thermal baths and operated as a two-stroke quantum thermal machine. Its working substance consists of a two-level system (TLS), generated by an unknown weak magnetic field acting on a qubit, and a second TLS arising due to the application of a known strong and tunable field on another qubit. Depending on this field, the machine may either act as an engine or a refrigerator. Under feasible conditions, determining this transition point allows to reduce the relative error of the measurement of the weak unknown magnetic field by the ratio of the temperatures of the colder bath to the hotter bath. 
\end{abstract}

\maketitle

\section{Introduction}
\label{sec_intro}

The recent experimental advances in the control of systems at the microscopic level enabled fascinating progress in quantum technologies~\cite{Nickerson14freely, kurizki15quantum, rossnagel16a, bernien17probing}. This has, in turn, sparked rigorous theoretical~\cite{paris09quantum, correa15individual, zwick16criticality, zhou18achieving} and experimental~\cite{kacprowicz10experimental, kucksko13, toyli13fluorescence} progress in the field of quantum metrology, with the aim of developing sensors capable of probing systems in the quantum regime with high accuracy~\cite{giovannetti11, degen17quantum, zwick17quantum, pezze18quantum}; such accuracy is essential throughout different branches of physics, including quantum information processing~\cite{wineland11quantum, li13entanglement, strobel14, toth14quantum}, quantum optics~\cite{dowling08quantum, joo11quantum} and condensed matter physics~\cite{giesbers08quantum,  zanardi08quantum, gross12spin, hovhannisyan18measuring}. One of the major challenges in quantum metrology is the precise estimation of small parameters~\cite{paris09quantum, brunelli11}. For example, high-precision low-temperature thermometry~\cite{kucksko13, correa15individual, pati19quantum} and high-precision magnetometry~\cite{muessel14scalable, brask15improved, albarelli17ultimate} have received a lot of attention in the quantum metrology community, owing to their immense importance in experimental realizations and applications.

\par

In parallel, it became crucial to understand and find the ultimate bounds of accuracy of parameter estimation~\cite{pezze09entanglement, dobrzanski12the}. The accuracy of estimating a parameter $x$ is quantified by the corresponding relative error $\hat{e}_x = \delta x / x$, where $\delta x$ denotes the absolute error of estimation. Previous studies have shown $\hat{e}_x$ to be lower-bounded by the Cramer-Rao bound, which in turn depends on the quantum Fisher information (QFI)~\cite{caves94, paris09quantum, zhang13, zwick16criticality, pasquale16local, gefen17}. In general, $\hat{e}_x$ increases as $x \to 0$. Consequently, 
developing ways of reducing the relative error of estimation of various parameters has been one of the major aims of the field of quantum metrology~\cite{zwick16criticality}. Recent studies have shown the possibility of using periodic control to enhance the precision of quantum probes~\cite{mukherjee17enhanced}, while other studies have suggested two-level systems with maximally degenerate excited states to be optimal for high-precision thermometry~\cite{correa15individual}. 

\par

In this work, we propose using a quantum thermal machine as a quantum probe. Quantum thermal machines are of great importance in the fields of quantum technologies and quantum thermodynamics~\cite{alicki1979quantum,scully03extracting, kosloff13quantum, zhang14quantum, klimovsky15thermodynamics, campisi16the, vinjanampathy16quantum,binder2019thermodynamicsbook}; at the same time, they were also shown to be beneficial for high-precision thermometry~\cite{brunner17quantum}. We present the possibility of using a quantum heat machine~\cite{uzdin15equivalence} as a magnetometer to estimate weak magnetic fields with high accuracy. To this end, we consider a pair ($\mathcal{K}$ and $\mathcal{U}$) of qubits (spin-1/2 particles), one ($\mathcal{K}$) subject to a {\it known} strong field, leading to a level splitting of $2\omegak$, and the second ($\mathcal{U}$) subject to a {\it unknown} weak magnetic field, resulting in a level splitting of $2\omegaun$, respectively. We aim to estimate $\omegaun$ by operating the above setup as a thermal machine whose cycle 
consists of two strokes~\cite{campisi2015nonequilibrium, campisi2016dissipation}. During the first, unitary stroke, the two TLSs are decoupled from the baths and allowed to interact with each other. During the second, thermalization stroke, $\mathcal{K}$ is allowed to thermally equilibrate with the hot thermal bath at temperature $\Th$ and $\mathcal{U}$ with a cold thermal bath at temperature $\Tc < \Th$. Depending on the known field, this thermal machine may either act as an engine or a refrigerator. The knowledge of $\omegak$, $\Th$ and $\Tc$ at the transition point between these two operation modes, i.e., the point of vanishing energy currents~\cite{klimovsky13minimal}, enables us to estimate the field $\omegaun$ with high accuracy.

\par

This paper is organized as follows. In Sec.~\ref{sec_setup} we introduce and discuss the setup of the two-stroke thermal machine used as 
a magnetometer. In Sec.~\ref{sec_swap} we discuss the operation of a thermal machine whose unitary stroke consists of swapping the populations of $\mathcal{K}$ and $\mathcal{U}$. 
An alternative machine whose unitary stroke generates entanglement between the two TLSs is discussed in Sec.~\ref{sec_mix}. The relative error of the magnetic field estimation and 
the corresponding QFI are investigated in Sec.~\ref{sec_err}.  In Sec.~\ref{sec_expt} we discuss the measurement of heat exchange between the hot bath and $\mathcal{K}$. Finally, 
we conclude in Sec.~\ref{sec_dis}.

\section{Model}
\label{sec_setup}

\par

\begin{figure}
  \includegraphics[width=0.95\columnwidth]{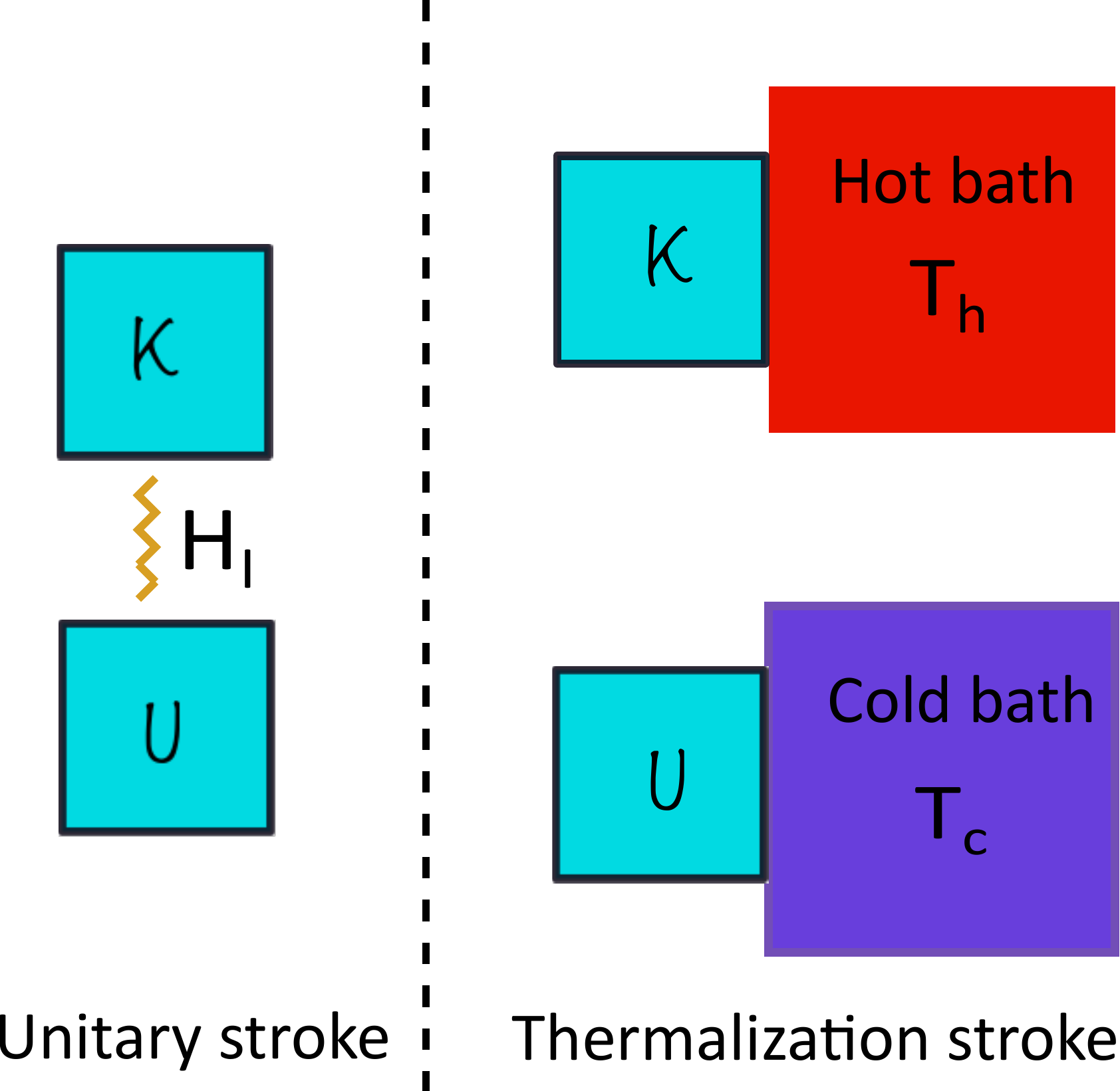}
  \caption{Schematic diagram of the two-stroke thermal machine (TTM) consisting of a two-level system (TLS) $\mathcal{K}$ with a known energy spectrum and a TLS $\mathcal{U}$ with unknown eigenenergies. During the unitary stroke $\lu$ both TLSs interact via the Hamiltonian $H_\mathrm{I}$ while during the thermalization stroke $\lt$ they independently equilibrate with the hot bath $\Th$ and cold bath $\Tc$, respectively.}\label{fig_TTM}
\end{figure}

\par

Let $\Hun = \omegaun\sigma_\mathrm{un}^z$ denote the local Hamiltonian of the TLS $\mathcal{U}$, where $\sigma_\mathrm{un}^z$ is the Pauli $z$-matrix, such that its energy eigenvalues are $\pm\omegaun$. The energy gap $2\omegaun$ is assumed to be generated by a weak unknown magnetic field and we strive to measure $\omegaun$ with high accuracy using the protocol described below. To this end we have at our disposal (see Fig.~\ref{fig_TTM}) another TLS $\mathcal{K}$, whose eigenspectrum is known and can be tuned by changing a parameter of the local Hamiltonian $\Hk = \omegak\sigma_\mathrm{k}^z$. We assume the known magnetic field that generates the gap {$2\omegak$} to be large, such that it can be measured using standard methods, for example, using superconducting quantum interference device~\cite{erne76theory, cleuziou06carbon}, with negligible relative error. The Hamiltonian of the total system then reads
\begin{equation}\label{eq_Htot}
  H(t) = \Hun + \Hk + H_\mathrm{I}(t) = \omegaun\sigma_\mathrm{un}^z + \omegak\sigma_\mathrm{k}^z + H_\mathrm{I}(t),
\end{equation}
where $H_\mathrm{I}(t)$ denotes the time-dependent interaction between the two TLSs when they are brought in contact. Additionally, we also have access to two (hot and cold) thermal baths maintained at temperatures $\Th$ and $\Tc<\Th$, respectively.

\par

Using the two TLSs and the baths $\Th$ and $\Tc$ allows us to construct a two-stroke thermal machine (TTM)~\cite{campisi2015nonequilibrium, campisi2016dissipation} as follows: Initially, the two TLSs $\mathcal{U}$ and $\mathcal{K}$ are isolated from each other, i.e., $H_\mathrm{I}(t)=0$, and in thermal equilibrium with the baths $\Tc$ and $\Th$, respectively. Their joint density operator is then the product state
\begin{equation}\label{eq_rho_tot}
  \rho_\mathrm{tot}(0)=\rho_\mathrm{k}(0)\otimes\rho_\mathrm{un}(0),
\end{equation}
where
\begin{subequations}\label{eq_denTLS}
  \begin{gather}
    \rho_\mathrm{un}(0)=
    \begin{pmatrix}   
      n_\mathrm{un} & 0\\
      0 & 1-n_\mathrm{un}
    \end{pmatrix}
    =\frac{1}{Z_\mathrm{un}}
    \begin{pmatrix}   
      e^{-{\frac{\omegaun}{\Tc}}} & 0\\
      0 & e^{{\frac{\omegaun}{\Tc}}}
    \end{pmatrix}\\
    \rho_\mathrm{k}(0)=\begin{pmatrix}   
      n_\mathrm{k} & 0\\
      0 & 1-n_\mathrm{k}
    \end{pmatrix}=\frac{1}{Z_\mathrm{k}}\begin{pmatrix}   
      e^{-{\frac{\omegak}{\Th}}} & 0\\
      0 & e^{{\frac{\omegak}{\Th}}}
    \end{pmatrix}.
  \end{gather}
\end{subequations} 
Here $n_\mathrm{un(k)}$ denotes the excited state population of $\mathcal{U}$~($\mathcal{K}$) and $Z_\mathrm{un(k)}=\Tr[\rho_\mathrm{un(k)}]$ are the respective partition functions. Note that we have set the Boltzmann and Planck constants to $k_\mathrm{B}=1$ and $\hbar=1$.

\par
  
On this initial configuration of the system we employ the following strokes (see Fig.~\ref{fig_TTM}):

\begin{enumerate}

\item Unitary Stroke ($\lu$): In this stroke, the TLSs are decoupled from the baths and the interaction $H_\mathrm{I}(t)$ is switched on at time $t=0$. Consequently, the two TLSs undergo a unitary evolution generated by the Hamiltonian \eqref{eq_Htot} for a duration $\tau_\mathrm{U}$. Thus, the final state of the combined system at time $t = \tau_\mathrm{U}$ depends on both $H_\mathrm{I}(t)$ and $\tau_\mathrm{U}$.

\item Thermalization Stroke ($\lt$): At time $t = \tau_\mathrm{U}$ the interaction $H_\mathrm{I}(t)$ is switched off and each TLS is again coupled to its respective thermal bath, i.e., the bath with which it was initially in equilibrium before the stroke $\mathcal{L}_\mathrm{U}$. The TLSs are kept in contact with the baths for a time $\tau_\mathrm{T}$ which is chosen to be large enough to allow the TLSs to thermalize back to their respective Gibbs states~\eqref{eq_denTLS}, i.e., the combined system returns to its initial configuration~\eqref{eq_rho_tot} and the next iteration of the cycle can start. We emphasize that heat exchanges between the TLSs and the baths only occur during this second stroke $\lt$.

\end{enumerate}

Let $\Qh$ and $\Qc$ denote the heat extracted from the baths $\Th$ and $\Tc$, respectively, during the stroke $\lt$. As the system returns to its initial configuration after the full cycle, energy is
conserved and the net work performed is therefore
\begin{equation}\label{eq_work}
  W=-(\Qh+\Qc).
\end{equation}
Depending on the sign of $\Qh$, $\Qc$ and $W$, TTMs may depict four distinct regimes of operation~\cite{buffoni2019quantum, mukherjee16speed},
  \begin{enumerate}
  \item Engine: $\Qh>0$, $\Qc<0$, $W<0$.	
  \item Refrigerator: $\Qh<0$, $\Qc>0$, $W>0$.
  \item Accelerator: $\Qh>0$, $\Qc<0$, $W>0$.	
  \item Heater: $\Qh<0$, $\Qc<0$, $W>0$.		
  \end{enumerate}
As we will see later, the TTMs considered here, however, can only operate in the first three of the above regimes. These regimes can be changed by appropriately tuning
the known strong magnetic field, i.e., tuning the Hamiltonian $\Hk$ of the TLS $\mathcal{K}$.

\par

Importantly, at the transition point between the engine and refrigerator operation modes, the work done as well as the heat exchanged between the system and the baths vanish, i.e., $\Qh=\Qc=W=0$. This is achieved when the excited state populations of the two TLSs [Eq.~\eqref{eq_denTLS}] coincide at the start of the cycle, i.e., $n_\mathrm{un}=n_{k}$. It is easy 
to see that this is satisfied if
\begin{equation}\label{eq_wun}
  \omegaun=\omegak^*\frac{\Tc}{\Th},
\end{equation}
where $\omegak^*$ denotes the energy eigenvalue of $\mathcal{K}$ at the transition point. In our protocol we repeat the cycle for different values of $\omegak$ (i.e., different strong magnetic fields), measure the energies $\Qh$, $\Qc$ and $W$, and identify the $\omegak^*$ at which the latter vanish; $\omegaun$ 
is then subsequently determined by Eq.~\eqref{eq_wun}.

\par

To motivate the use of our protocol we consider the relative error $\Delta \omegaun/\omegaun$. Since each of the quantities on the r.h.s of Eq.~\eqref{eq_wun} can be measured independently, 
the root mean square (rms) error $\Delta \omegaun$ in measuring $\w_\mathrm{un}$ can be obtained from the rms error propagation law,
\begin{widetext}
\begin{align}\label{eq_Delta_wun}
  \Delta \omegaun &= \sqrt{\left(\frac{\partial \omegaun}{\partial \w_\mathrm{k}^*}\right)^2(\Delta \w_\mathrm{k}^*)^2+\left(\frac{\partial \omegaun}{\partial \Th}\right)^2(\Delta \Th)^2+\left(\frac{\partial \omegaun}{\partial \Tc}\right)^2(\Delta \Tc)^2}\nonumber\\
  &=\sqrt{\left(\frac{\Tc}{\Th}\right)^2(\Delta \w_\mathrm{k}^*)^2+\left(\frac{\w_\mathrm{k}^*\Tc}{\Th^2}\right)^2(\Delta \Th)^2+\left(\frac{\w_\mathrm{k}^*}{\Th}\right)^2(\Delta \Tc)^2},
\end{align}
\end{widetext}
such that the relative error is
\begin{align}\label{eq_rel_Delta_wun}
\frac{\Delta \omegaun}{\omegaun} =\sqrt{\left(\frac{\Delta \w_\mathrm{k}^*}{\w_\mathrm{k}^*}\right)^2+\left(\frac{\Delta \Th}{\Th}\right)^2+\left(\frac{\Delta \Tc}{\Tc}\right)^2},
\end{align}
where $\Delta x$ denotes the rms error in measuring the quantity $x$. This treatment requires $\omegaun$ to be unimodal (see Appendix~\ref{app_unimodal}). We now assume that we model
the setup using thermal baths whose temperatures are known to a high degree of accuracy. This should be possible in light of the recent developments in high-precision low-temperature thermometry which
have enabled experimentalists to measure temperatures as low as $6$ mK, with an accuracy of $0.1$ mK~\cite{iftikhar16primary}. Furthermore, the relative error bound of thermometry
decreases as $1/\sqrt{\mathcal{M}_T}$ for $\mathcal{M}_T$ being the number of temperature measurements~\cite{paris09quantum}. Consequently, one can always expect to know the bath temperatures with negligibly
small relative error (i.e., $\Delta \Th/\Th \approx \Delta \Tc/\Tc\approx 0$) by considering large enough $\mathcal{M}$, such that Eqs.~\eqref{eq_wun} and~\eqref{eq_rel_Delta_wun} yield
\begin{equation}\label{eq_dwun}
  \Delta \omegaun=\Delta\omegak^*\frac{\Tc}{\Th}.
\end{equation}
On a related note, recent experiments on microscopic heat machines have modelled baths using electric-field noise \cite{rossnagel16a, klaers17squeezed}.
\par

Note that determining $\omegak^*$ involves two kinds of measurements. A first measurement is required to identify the transition point via the (vanishing) quantities $\Qh$, $\Qc$ and $W$ 
followed by a direct measurement of $\omegak^*$ once the transition point is reached. Each of these measurements can be carried out independently such that the rms error $\Delta\omegak^*$ evaluates to
\begin{equation}\label{eq_dwun2}
  \Delta\omegak^* = \sqrt{ (\Delta\w'_\mathrm{k})^2 + \sum_i\alpha_i^2(\Delta Q_i)^2},
\end{equation}
where $\alpha_i=\left|\frac{\partial{\omegak}}{\partial{Q_i}}\right|_{\omegak^*}$ with $Q_1=\Qh$, $Q_2=\Qc$ and $Q_3=W$. Here, the first term $\Delta\w'_\mathrm{k}$ is the error in directly measuring 
$\omegak$. While this error stems from the experimental apparatus used for measuring $\omegak$, the remaining terms in Eq.~\eqref{eq_dwun2} arise from the respective errors in measuring the energy 
currents $Q_i$. Therefore, we can equate $\Delta\w'_\mathrm{k}$  with the error that one would have obtained if the TLS $\mathcal{U}$ was also measured directly, i.e., without using our protocol. 
Substituting  $\Delta\w'_\mathrm{k} \approx \Delta\w'_\mathrm{un}$ and plugging Eq.~\eqref{eq_dwun2} in Eq.~\eqref{eq_dwun} and dividing by $\omegaun$, we obtain
\begin{equation}\label{eq_dwk}
\frac{\Delta\omegaun}{\omegaun} = \frac{\Tc}{\Th}\sqrt{ \left(\frac{\Delta\w'_\mathrm{un}}{\omegaun}\right)^2 + \sum_i\frac{\alpha_i^2(\Delta Q_i)^2}{\omega_\mathrm{un}^2}}.
\end{equation}

\par

We now further assume that $\sum_i\alpha_\mathrm{i}^2(\Delta Q_i)^2\ll(\Delta\w'_\mathrm{un})^2$. The first requirement for this assumption to hold true is that the coefficients 
$\alpha_i$ should not diverge to a large value. As we will show later, for our TTM $\alpha_i\sim\mathcal{O}(1)$ such that this requirement is met. Secondly, the measurement of $Q_i$ 
must be precise enough such that the error $\Delta Q_i$ is sufficiently small. The advantage of our
proposed protocol is that one only needs to  look for a sign-reversal in the quantity $Q_i$, which in general can be more precise than measuring the absolute
value of the quantity $Q_i$. Therefore, we emphasize here that 
even though any of the quantities $\Qh$, $\Qc$ or $W$ can be measured to identify $\w_\mathrm{k}^*$, the most favourable choice is to measure the quantity that undergoes a sign reversal.
In our case, as we will demonstrate later, the sign reversal is guaranteed in $\Qh$ and $\Qc$ but not in $W$. However, we note that the accurate and precise measurement of heat exchanges remains a highly challenging task under currently available experimental resources. Nevertheless, one can resort to obtaining $\Qh$ using indirect ways that are in general much more precise than direct heat measurements; for example, by measuring the magnetization or, equivalently, the state of the TLS $\mathcal{K}$ (see Sec.~\ref{sec_expt}). Hence, we focus on the regime where the second term on the r.h.s of Eq.~\eqref{eq_dwk} is not only small but can also be neglected under appropriate experimental conditions. Consequently, the error reduces to
\begin{equation}\label{eq_dwun3}
  \Delta \omegaun\approx\Delta\w'_\mathrm{un}\frac{\Tc}{\Th}. 
\end{equation}
 This equation reveals the important result that any error arising in the direct measurement of $\omegaun$ (or equivalently the error in measuring $\omegak$) is scaled down by a factor of $\Tc/\Th$ and therefore can be reduced if $\Th$ is chosen sufficiently large. Our technique is therefore reminiscent of the Wheatstone bridge setup where the error in measuring small electrical resistances can be significantly reduced through an indirect measurement of a higher resistance.  A similar technique of using a quantum thermal machine as a thermometer is also presented in Ref.~\cite{brunner17quantum}. Further, such enhancements in precision using the Wheatstone bridge principle have also been demonstrated in a number of other measurement devices, e.g., for enhancing sensitivity of  thermal conductivity measurements of nanostructures~\cite{wingert2012ultra}. In what follows, we illustrate this point in more 
detail by considering two particular examples of $H_\mathrm{I}(t)$, i.e., two protocols realizing the TTM in Fig.~\ref{fig_TTM}.

\section{Two-stroke thermal machines as quantum probes}

\subsection{Swap TTM}
\label{sec_swap}

We first consider an interaction $H_\mathrm{I}(t)$ that swaps the states of $\mathcal{U}$ and $\mathcal{K}$ at the end of the unitary stroke $\mathcal{L}_\mathrm{U}$, such that the total density operator remains a product state~\cite{sangouard05fast, liang05realization, zhang07mutual, uzdin15equivalence}. This protocol realizes the \textit{swap TTM} and at the end of stroke $\mathcal{L}_\mathrm{U}$ the states of the two TLSs read
\begin{subequations}\label{eq_denTLS2}
  \begin{gather}
    \rho_\mathrm{un}(\tau_{U})=\begin{pmatrix}   
      n_\mathrm{k} & 0\\
      0 & 1-n_\mathrm{k}
    \end{pmatrix}=\frac{1}{Z_\mathrm{k}}\begin{pmatrix}   
      e^{-{\frac{\omegak}{\Th}}} & 0\\
      0 & e^{{\frac{\omegak}{\Th}}}
    \end{pmatrix}\\
    \rho_\mathrm{k}(\tau_{U})=\begin{pmatrix}   
      n_\mathrm{un} & 0\\
      0 & 1-n_\mathrm{un}
    \end{pmatrix}=\frac{1}{Z_\mathrm{un}}\begin{pmatrix}   
      e^{-{\frac{\omegaun}{\Tc}}} & 0\\
      0 & e^{{\frac{\omegaun}{\Tc}}}
    \end{pmatrix}.
  \end{gather}
\end{subequations} 
After the thermalization stroke $\lt$ the system returns to its initial configuration~\eqref{eq_rho_tot},
\begin{equation}
  \rho_\mathrm{un(k)}(\tau_\mathrm{U}+\tau_\mathrm{T})=\rho_\mathrm{un(k)}(0).
\end{equation}

\par

We recall here that the TLS $\mathcal{U}$ is isolated from $\mathcal{K}$ and kept in contact with bath $\Tc$ during the stroke $\lt$. The heat $\Qc$ extracted from the bath $\Tc$ therefore equals the change in the internal energy of $\mathcal{U}$,
\begin{subequations}\label{eq_swap_heat}
  \begin{multline}\label{eq_swap_heat1}
    \Qc={\Tr}[\rho_\mathrm{un}(\tau_{U}+\tau_\mathrm{T})\Hun] - {\Tr}[\rho_\mathrm{un}(\tau_{U})\Hun]\\
    =2\omegaun(n_\mathrm{un}-n_\mathrm{k}).
  \end{multline}
  Similarly, the heat $\Qh$ extracted by $\mathcal{K}$ from $\Th$ is
  \begin{equation}\label{eq_swap_heat2}
    \Qh=2\omegak(n_\mathrm{k}-n_\mathrm{un}),
  \end{equation}
  such that, using Eq.~\eqref{eq_work}, the performed work is
  \begin{equation}
    W=-2(\omegak-\omegaun)(n_\mathrm{k}-n_\mathrm{un}).
  \end{equation}
\end{subequations}

\par

From Eqs.~\eqref{eq_swap_heat}{, it} follows that $\Qc=\Qh=W=0$ when $n_\mathrm{un}=n_\mathrm{k}$. This condition is satisfied when $\omegak^*=\omegaun\Th/\Tc$ [Eq.~\eqref{eq_wun}]. Now, consider $\omegak^+=\omegak^* + \epsilon$ with $\epsilon>0$. It immediately follows that $n_\mathrm{k}<n_\mathrm{un}$. In addition, since $\Th>\Tc$, we find
\begin{equation}
  \omegak^+ =\omegaun\left(\frac{\Th}{\Tc} + \frac{\epsilon}{\omegaun}\right)>\omegaun 
\end{equation}
and hence $\Qh<0$ and $W, \Qc>0$. The TTM therefore works as a quantum refrigerator.

\par

On the other hand, considering $\omegak^-=\omegak^* - \epsilon$, we have $n_\mathrm{k}>n_\mathrm{un}$ and
\begin{equation}
  \omegak^- =\omegaun\left(\frac{\Th}{\Tc} - \frac{\epsilon}{\omegaun}\right).
\end{equation}
Two possible modes of operation are now possible: For $\epsilon<\omegaun(\Th/\Tc-1)$, we have $\omegak^->\omegaun$. Consequently, $\Qh>0$, $\Qc<0$ and $W<0$ --- the TTM hence works as a
quantum heat engine. On the other hand, if $\epsilon>\omegaun(\Th/\Tc-1)$, we get $\omegak^-<\omegaun$. Hence, $\Qh>0$, $\Qc<0$ and $W>0$, which corresponds to an accelerator where work is
performed to amplify the transfer of heat from the hot to the cold bath. Thus, if one starts from a high $\omegak$ ($>\omegak^*$) and progressively lowers its value, a transition from  QR to
QHE occurs at $\omegak^*$, followed by a transition to the accelerator regime. On a related note, the above QHE-QR transition has also been shown in four-stroke~\cite{kosloff17the} and
 continuous thermal machines~\cite{klimovsky13minimal, kosloff14quantum, mukherjee16speed}. Note that,
while $\Qh$ and $\Qc$ reverse their sign only at the first transition (QR-QHE), $W$ undergoes sign reversal at the QHE-accelerator transition as well.
Therefore, $\Qh$ and $\Qc$ are the proper quantities of measurement for detecting the QHE-QR transition  in our protocol.

\par

Having established the existence of a transition point for the swap TTM, we now proceed to make an estimate of the error in estimating $\omegaun$. The coefficients $\ah$ and $\ac$ defined in Eq.~\eqref{eq_dwun2} evaluate to
\begin{subequations}\label{eq_der_qh}
  \begin{gather}
    \ah^{-1}=\left|\frac{\partial{Q_\mathrm{h}}}{\partial{\omegak}}\right|_{\omegak^*}=\left(\frac{\omegaun}{\Tc}\right)\sech^2{\left(\frac{\omegaun}{\Tc}\right)}\label{eq_der_qh1}\\
    \ac^{-1}=\left|\frac{\partial{Q_\mathrm{c}}}{\partial{\omegak}}\right|_{\omegak^*}=\left(\frac{\omegaun}{\Th}\right)\sech^2{\left(\frac{\omegaun}{\Tc}\right)}=\left(\frac{\Tc}{\Th}\right)\ah^{-1}.\label{eq_der_qc1}
  \end{gather}
\end{subequations}
As explained above, measuring any one of the quantities $\Qh$ or $\Qc$ is sufficient to identify the transition point. Further, Eq.~\eqref{eq_der_qc1} reveals that $\ac>\ah$, and therefore we conclude that $\Qh$ is the preferred quantity of measurement for detecting the transition point since $\omegak$ is less sensitive to experimental errors in measuring $\Qh$ as compared to $\Qc$. We will show later that the quantity $\ah$ is closely related to the quantum Fisher information (QFI) for a TLS initialized in thermal equilibrium with a bath.
\begin{figure*}
	\centering
	\begin{center}
		\subfigure{
			\includegraphics[width=0.45\textwidth]{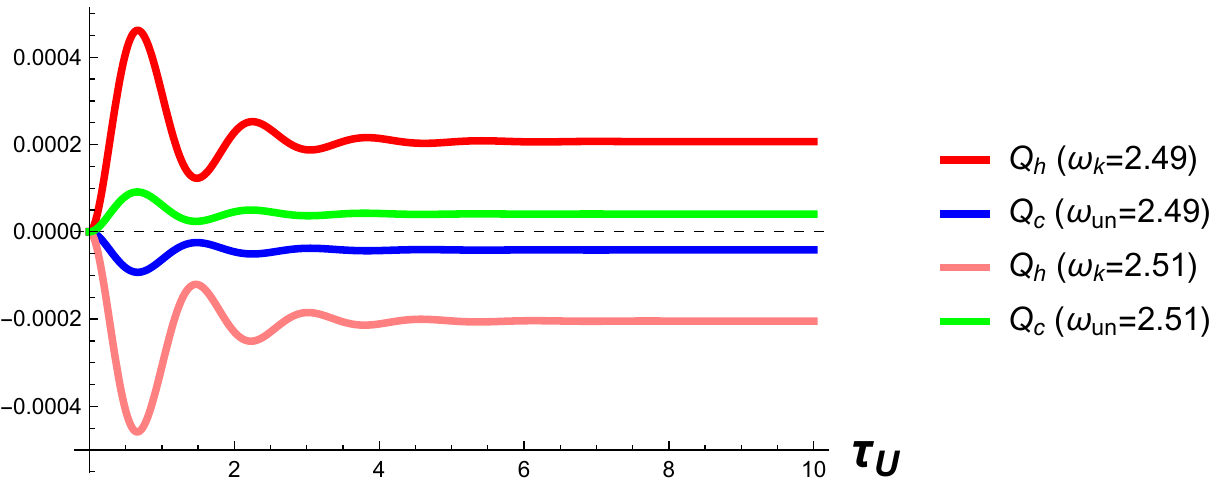}
			\label{fig1}
		}
		\subfigure{
			\includegraphics[width=0.45\textwidth]{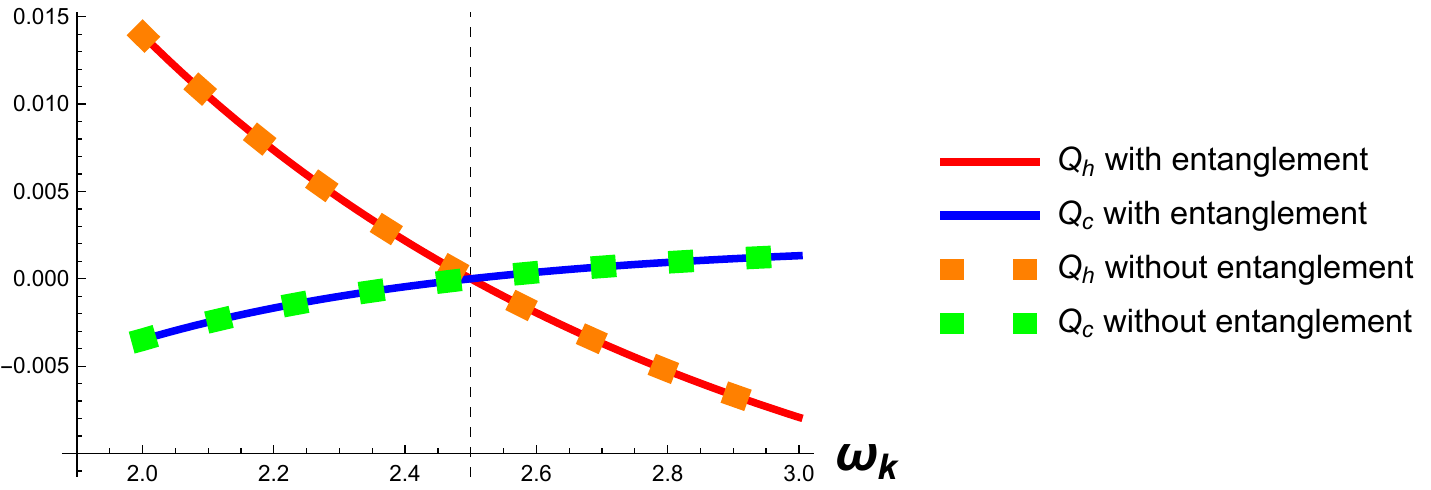}
			\label{fig2}
		}
	\end{center}
	\caption{Heat~\eqref{eq_mix_heat} exchanged with the two baths (left panel) as a function of the unitary stroke duration $\tau_\mathrm{U}$ for $\omegak$ both above and below the transition 
	point and (right panel) as a function of $\omegak$ for $\tau_\mathrm{U}=10$. The transition point is therefore $\omegak^*=2.5$. The right panel also shows the equivalence of
	the heat exchanges calculated using Eq.~\eqref{eq_mix_heat} 
	(dashed lines) to those calculated using a master equation approach (Eq.~\eqref{eq_QE} in Appendix.~\ref{app_heat}) (solid lines). The other parameters are $a=1$, $\omegaun=0.5$, $\w_\mathrm{I}=1$, $\Tc=1$, $\Th=5$, $\tau_\mathrm{T}=10$, $\tau_\mathrm{U}=10$, $\gamma_\mathrm{k}=1$ and $\gamma_\mathrm{un}=1$.}
\end{figure*}

\subsection{Mix TTM}
\label{sec_mix}

We now study another TTM, henceforth referred to as \textit{mix TTM}, with a more generic interaction $H_\mathrm{I}(t)$ than the swapping operation considered in Sec.~\ref{sec_swap} and in which the existence of a transition point is not a priori apparent. Contrary to the swap TTM, in the mix TTM the unitary stroke $\lu$ may generate an entangled state of $\mathcal{K}$ and $\mathcal{U}$.

\par

We choose the interaction to be of the form
\begin{equation}\label{eq_HI}
  H_\mathrm{I}(t)= \w_\mathrm{I}f(t)\left(\sigma_\mathrm{k}^x\sigma_\mathrm{un}^x+\sigma_\mathrm{k}^y\sigma_\mathrm{un}^y\right),
\end{equation}
where $f(t)$ encodes the time dependence and $\w_{I}$ is the characteristic interaction energy scale. Contrary to the swap TTM from Sec.~\ref{sec_swap}, only a partial swap of 
the TLSs population occurs at the end of the stroke $\lu$ and the combined system remains in an entangled state.  As shown in Appendix~\ref{app_heat}, the heat exchanges can be evaluated as
\begin{equation}\label{eq_mix_heat}
  Q_\mathrm{h(c)}={\Tr}[\rho_\mathrm{k(un)}(\tau_\mathrm{U}+\tau_\mathrm{T})H_\mathrm{k(un)}] - {\Tr}[\rho_\mathrm{k(un)}(\tau_\mathrm{U})H_\mathrm{k(un)}]
\end{equation}
with 
\begin{equation}\label{eq_rku_mix}
  \rho_\mathrm{k(un)}(\tau_\mathrm{U})={\Tr}_{\mathrm{un}(\mathrm{k})}[\rt(\tau_\mathrm{U})].
\end{equation}

\par

One can in principle derive an exact analytical expression for Eq.~\eqref{eq_mix_heat} to investigate the presence of a transition point. However, we argue that the heat exchanged and the work done are guaranteed to vanish at $n_\mathrm{k}=n_\mathrm{un}$ due to the particular form of $H_\mathrm{I}(t)$ that we choose for this TTM. To elaborate, we rewrite $H_\mathrm{I}(t)$ from Eq.~\eqref{eq_HI} as
\begin{equation}
  H_\mathrm{I}(t) = 2\w_\mathrm{I} f(t)\left(\ket{\uparrow_\mathrm{k}}\ket{\downarrow_\mathrm{un}}\bra{\downarrow_\mathrm{k}}\bra{\uparrow_\mathrm{un}} + \mathrm{h.c.} \right),
\end{equation}
where \{$\ket{\uparrow_\mathrm{k(un)}}\ket{\downarrow_\mathrm{k(un)}}$\} is the eigenbasis of $\sigma_\mathrm{k(un)}^z$. The unitary evolution of $\rt(0)$ under the action of the above $H_\mathrm{I}(t)$ during the unitary stroke $\lu$ therefore only rotates the projection of $\rt(0)$ on the subspace spanned by the states $\ket{\uparrow_\mathrm{k}}\ket{\downarrow_\mathrm{un}}$ and $\ket{\downarrow_\mathrm{k}}\ket{\uparrow_\mathrm{un}}$. 
For an arbitrary (non-zero) rotation of $\theta$ on this subspace, the heat exchanged and the work done can be expressed (see Appendix~\ref{app_theta}) as 
\begin{subequations}
  \begin{align}
    \Qc&=2\omegaun(n_\mathrm{un}-n_\mathrm{k})\sin^2{\theta}\\
    \Qh&=2\omegak(n_\mathrm{k}-n_\mathrm{un})\sin^2{\theta}\label{eq_mix_heat2}\\
    W&=-2(\omegak-\omegaun)(n_\mathrm{k}-n_\mathrm{un})\sin^2{\theta}.
  \end{align}
\end{subequations} 
Similarly to Sec.~\ref{sec_swap}, one can accordingly categorize the three regimes of operation (engine, refrigerator and accelerator). In particular, a QHE-QR transition occurring 
at $n_\mathrm{k}=n_\mathrm{un}$ is found as the point where all energy currents vanish [see Fig.~\ref{fig2}]. Further, since $W$ reverses sign at both the QR-QHE  and QHE-accelerator 
transitions, we look for a sign reversal in either $\Qh$ or $\Qc$ to detect the former transition which is the relevant one for our protocol. Note that for $\theta=\pi/2$ the energy 
currents~\eqref{eq_swap_heat} of the swap TTM are recovered.

\par

Note that while any interaction of the form~\eqref{eq_HI} guarantees the vanishing of the energy currents at $n_\mathrm{k}=n_\mathrm{un}$, only a judicious choice of the time dependence $f(t)$ eliminates any dependence of $Q_\mathrm{h(c)}$ on the duration $\tau_\mathrm{U}$ of the unitary stroke $\lu$. 
In  Fig.~\ref{fig1} , we show that for an exponentially-decaying interaction, i.e., $f(t)=e^{-at}$ with $a>0$, the heats 
exchanged with the baths $Q_\mathrm{h(c)}$ become independent of $\tau_\mathrm{U}$ if $\tau_\mathrm{U}\gg(1/a)$. This suggests that one does not need to maintain the same $\tau_\mathrm{U}$ while 
performing repeated measurements as long as $\tau_\mathrm{U}\gg(1/a)$ is satisfied.

\par

While the duration $\tau_{U}$ of the stroke $\lu$ has to be much greater than the timescale of the interaction between the two TLSs, the duration $\tau_{T}$ of the stroke $\lt$, on the 
other hand, needs to be sufficiently larger than the thermalization timescale $\tau_\mathrm{th}$ on which the TLS thermalize to their respective Gibbs states. The latter is determined by
$\tau_\mathrm{th}=\min\{|\realp[\lambda_i]|\neq 0\}^{-1}=\gamma^{-1}(1+e^{-2\w_\mathrm{un}/\Tc})^{-1}$, where $\lambda_i$ are the eigenvalues of the propagator for the stroke $\lt$ and the decay 
rate $\gamma$ is proportional to the square of the coupling strength between the TLSs and the bath~\cite{breuerbook}. For the purpose of simplicity, we have here assumed equal $\gamma$ for both 
TLSs (see Appendix~\ref{app_therm}). As we will show later in Sec.~\ref{sec_err}, for our protocol to work efficiently the ratio $\w_\mathrm{un}/\Tc$ should be close to unity ($\approx 1.2$). 
Consequently, we have $\tau_\mathrm{th}\propto\gamma^{-1}$. However, note that while one may be tempted to choose a high value of $\gamma$ to have a small $\tau_\mathrm{th}$, in case 
of the thermalising strokes being
described by Markovian dynamics, the secular approximation~\cite{breuerbook} requires $\tau_\mathrm{th}$ to be much larger than the intrinsic timescale of the TLSs, which in our case is of the order 
of $1/(2\omegaun)$. 
The necessary condition, under secular approximation, is therefore $\tau_{T}\gg1/(2\omegaun)$, which means that estimating very low magnetic fields requires longer cycle times. 
However, the magnetometry protocol proposed here can be expected to be beneficial as long as the engine-to-refrigerator transition (or vice-versa), i.e., vanishing $\Qh$, $\Qc$ and $W$,
occurs under the  condition~\eqref{eq_wun}. Furthermore, the values of $\Qh$, $\Qc$, and hence $W$, depend on only the initial and final states of the two strokes, rather than on the details of
the strokes. Consequently, the constraints on the timescales may be relaxed for non-unitary strokes which do not rely on secular approximation.

\par

\begin{figure}
  \centering
  \includegraphics[width=0.45\textwidth]{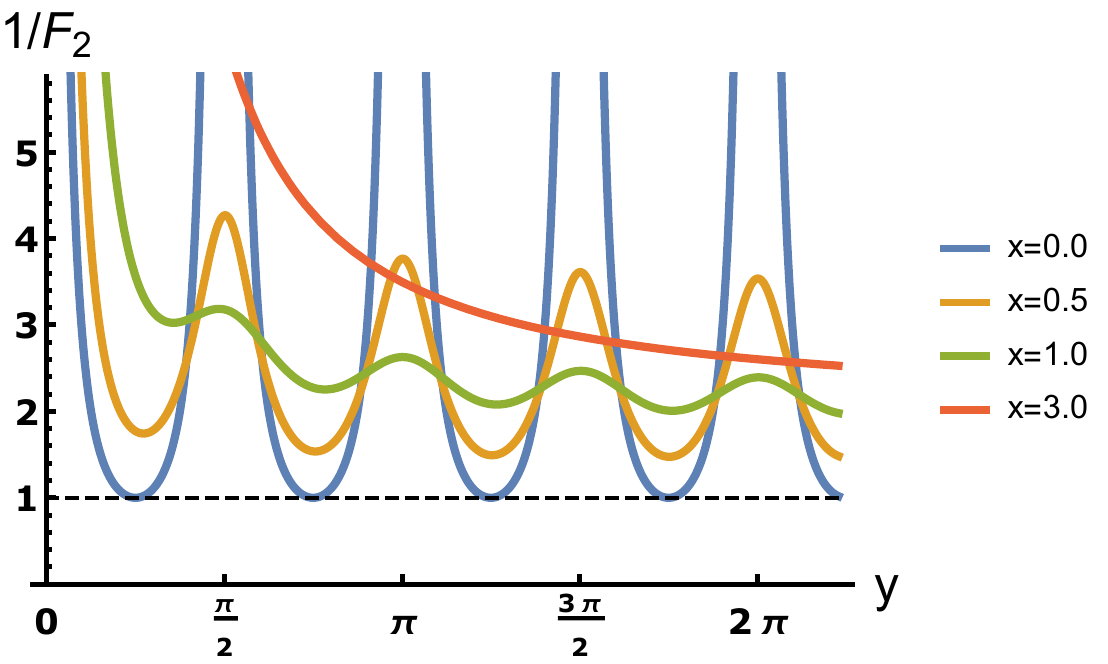}
  \caption{Function $1/F_2(x,y)$ plotted against $y=\w_\mathrm{I}/a$ for different different values of $x=(\omegaun-\omegak^*)/a$. In the limit $x\to 0$ ($a\gg \omegaun-\omegak^*$), $1/F_2$ approaches the minimum value of unity (dashed) at $y=(2n+1)\pi/4$, where $n$ is a positive integer.}\label{fig_f2}
\end{figure}

Using Eq.~\eqref{eq_mix_heat}, the coefficients $\ac$ and $\ah$ in Eq.~\eqref{eq_dwun2} for the error $\Delta\omegak^*$ evaluate to
\begin{subequations}
  \begin{gather}\label{eq_der_h2}
    \ah^{-1}=\left|\frac{\partial{Q_\mathrm{h}}}{\partial{\omegak}}\right|_{\omegak^*}=F_1\left(\frac{\omegaun}{\Tc}\right)F_2\left(\frac{1}{a}(\omegaun-\omegak^*), \frac{\w_\mathrm{I}}{a}\right)\\
    \ac^{-1}=\left|\frac{\partial{Q_\mathrm{c}}}{\partial{\omegak}}\right|_{\omegak^*}=\left(\frac{\Tc}{\Th}\right)\ah^{-1},
  \end{gather}
\end{subequations}
where we defined
\begin{subequations}
  \begin{equation}
    F_1(x)=x\sech^2{\left(x\right)}
  \end{equation}
  and
  \begin{multline}
    F_2(x,y)=\pi^2y^2\sech^2\left(\pi x\right)\left|J_{\frac{1}{2}+ix}\left(2y\right)\right|^2\\\times\left(\lim_{t\to\infty}\left|J_{-\frac{1}{2}+ix}\left(2e^{-at}y\right)\right|^2e^{-at}\right),
  \end{multline}
\end{subequations}
$J_z(s)$ being the Bessel function of first kind. Note that $F_1(x)$ already appeared in Eq.~\eqref{eq_der_qh1} for the swap TTM and, as we will show later, is related to the QFI. Fig.~\ref{fig_f2} shows that in the limit $\omegaun-\omegak^*\ll a$, $1/F_2$ attains a minimum value of unity at $\w_\mathrm{I}/a=(2n+1)\pi/4$, where $n$ is a positive integer. Therefore, the minimum value of $\alpha_\mathrm{h}$ for the mix TTM matches its counterpart~\eqref{eq_der_qh1} for the swap TTM. Note that the additional factor $F_2(x)$ is an artefact stemming from the fact that in contrast to the swap TTM the excited state population of the TLSs are not completely interchanged in the mix TTM.

\par

It is interesting to note that the transition point discussed in the manuscript is actually the \textit{Carnot} point at which the TTMs achieve reversibility. It can be shown that at the
transition point the unitary stroke effectively becomes an identity transformation of the system density matrix, which may be considered an adiabatic transformation. Similarly, the system remains 
in thermal equilibrium with the baths during the thermalization stroke and hence the latter may be seen as an isothermal process. Therefore, the two stroke cycle becomes reversible at the transition point. When working as an engine, the efficiency $\eta=|W|/\Qh$ of the discussed TTMs evaluates to

  \begin{equation}
  \eta = 1-\frac{\Qc}{\Qh}=1-\frac{\omegaun}{\omegak},
\end{equation}

Using Eq.~\eqref{eq_wun}, one can check that the efficiency converges towards the Carnot efficiency as the transition point is approached,
\begin{equation}
  \lim_{\omegak \to\omegak^*}\eta=1-\frac{\omegaun}{\omegak^*}=1-\frac{\Tc}{\Th}.
\end{equation}
This behaviour has also been previously reported for the Otto cycle~\cite{kosloff17the} and continuous thermal machines~\cite{klimovsky13minimal, klimovsky15thermodynamics}.

\section{Minimal achievable relative error}
\label{sec_err}

We now address the relative error in measuring $\omegaun$ using the two TTMs discussed above. We reiterate that we only measure $\Qh$ for determining the transition point.
Equation~\eqref{eq_dwk} then yields
\begin{align}\label{eq_err_rel_1}
  \frac{\Delta \omegaun}{\omegaun}&=\sqrt{\left(\frac{\Tc}{\Th}\right)^2\left(\frac{\Delta \omegaun'}{\omegaun}\right)^2+\ah^2\left(\frac{\Tc}{\Th}\right)^2\frac{(\Delta \Qh)^2}{\omegaun^2}}\notag\\
  &=\sqrt{\left(\frac{\Tc}{\Th}\right)^2\left(\frac{\Delta \omegaun'}{\omegaun}\right)^2+F^2\bar{\alpha}_\mathrm{h}^2\frac{(\Delta \Qh)^2}{\Th^2}},
\end{align}
where we have defined
\begin{equation}\label{eq_alphabar}
  \bar{\alpha}_\mathrm{h}=\left[\frac{\omegaun}{\Tc}\sech{\left(\frac{\omegaun}{\Tc}\right)}\right]^{-2}.
\end{equation}
Further, $F=1$ for the swap TTM and $F=1/F_2((\omegaun-\omegak^*)/a, \w_\mathrm{I}/a)$ for the mix TTM; note that the latter can be minimized to unity 
(see Fig.~\ref{fig_f2}). Even though the proposed protocol would allow us to measure $\omegaun$ as long as Eq. \eqref{eq_wun} holds, 
however, the magnetic field estimation would be most accurate in the limit 
\begin{align}\label{eq_err_con}
 \bar{\alpha}_\mathrm{h}^2\frac{(\Delta \Qh)^2}{\Th^2} \ll   \left(\frac{\Tc}{\Th}\right)^2\left(\frac{\Delta \omegaun'}{\omegaun}\right)^2.
\end{align}
	
\par

\begin{figure}
  \centering
  \includegraphics[width=0.45\textwidth]{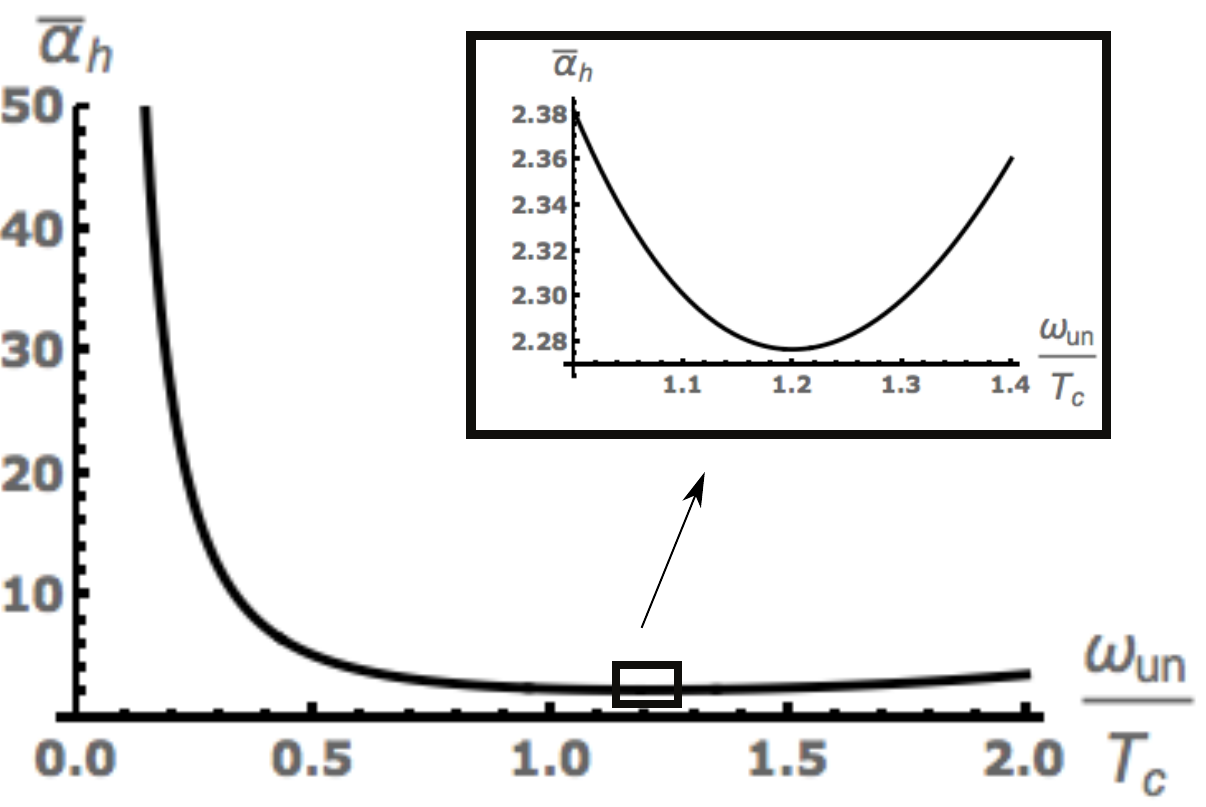}
  \caption{Variation of $\bar{\alpha}_\mathrm{h}$ [Eq.~\eqref{eq_alphabar}] with $\omegaun/\Tc$. The inset shows that $\bar{\alpha}_\mathrm{h}$ is minimized for $\omegaun/\Tc\approx 1.2$. This sets the condition for choosing an optimum value of the cold bath $\Tc$ from a rough estimate of $\omegaun$.}\label{fig_alpha}
\end{figure}

\par
Fig.~\ref{fig_alpha} shows that the quantity $\bar{\alpha}_\mathrm{h}$ [Eq.~\eqref{eq_alphabar}] is minimized when $\omegaun/\Tc\approx 1.2$
($\bar{\alpha}_\mathrm{h}\approx 2.28$, $\alpha_h\sim\mathcal{O}(1)$). At these optimum values of $\bar{\alpha}_\mathrm{h}$ and $\omegaun/\Tc$, the above condition reduces to 
\begin{align}\label{eq_err_con2}
(\Delta \Qh)^2 \ll  \frac{(\Delta \omegaun')^2}{10} .
\end{align}
Our protocol therefore requires a precise detection (zero measurement) of $\Qh$ at the transition point, which can, for example, be done through a measurement of the state of the qubit $\mathcal{K}$, as discussed in Sec. \ref{sec_expt} below.

\par

It is interesting to note that the quantity $\bar{\alpha}_\mathrm{h}$ is intricately connected to the quantum Fisher information (QFI) corresponding to a TLS initialized in thermal equilibrium with a bath. To elucidate this connection, we recall that the quantum version of the Cramer-Rao bound dictates that the minimal achievable relative error is bounded by the inequality~\cite{caves94, paris09quantum}
\begin{equation}\label{eq_crbound}
  \frac{\Delta\omegaun}{\omegaun}\geq\frac{1}{\omegaun\sqrt{\mathcal{F}_\mathrm{I}(\omegaun)}},
\end{equation}%
where
\begin{equation}\label{eq_qfi}
  \mathcal{F}_\mathrm{I}(\omegaun)=-2\lim_{\epsilon\to 0}\frac{\partial^2}{\partial\epsilon^2}\mathcal{F}\Big(\rho_\mathrm{un}(\omegaun+\epsilon),\rho_\mathrm{un}(\omegaun)\Big)
\end{equation}
is the quantum Fisher information (QFI) and $\mathcal{F}(\rho_1,\rho_2)=\Tr[\sqrt{\sqrt{\rho_1}\rho_2\sqrt{\rho_1}}]$ is the fidelity between the states $\rho_1$ and $\rho_2$~\cite{paris09quantum}.

\par

The QFI for the TLS $\mathcal{U}$ prepared in thermal equilibrium with the bath $\Tc$ assumes the simple form (see Appendix~\ref{app_QFI})
\begin{equation}	
  \mathcal{F}_\mathrm{I}(\omegaun)=\frac{1}{\Tc^2}\sech^2{\left(\frac{\omegaun}{\Tc}\right)},
\end{equation}
such that Eq.~\eqref{eq_crbound} evaluates to
\begin{equation}\label{eq_relerr}
  \frac{\Delta\omegaun}{\omegaun}\geq\left[\frac{\omegaun}{\Tc}\sech{\left(\frac{\omegaun}{\Tc}\right)}\right]^{-1}=\sqrt{\bar{\alpha}_\mathrm{h}}.
\end{equation}

\par

The above inequality leads us to conclude the following: The relative error in the measurement of $\omegaun$ has the theoretical lower bound~\eqref{eq_crbound} for any protocol that requires the TLS $\mathcal{U}$ to be initialized in thermal equilibrium with a bath at temperature $\Tc$. Since the quantum Cramer-Rao bound has a geometric origin in the density matrix space, its value can only be lowered by choosing optimum values of parameters that characterize the density matrix of the system. Given a TLS with a certain unknown $\omegaun$, the only free parameter that characterizes the state $\rho_\mathrm{un}(0)$ is the bath temperature $\Tc$. The optimization condition therefore, as can be seen from Eq.~\eqref{eq_relerr}, corresponds to choosing the bath temperature $\Tc$ such that the quantity $\bar{\alpha}_\mathrm{h}$ is minimized. It is intriguing to recall here that we encountered the same minimization requirement in our protocol through Eq.~\eqref{eq_err_con}, provided the temperature of the baths are accurately known.

\section{Indirect measurement of the heat $Q_\mathrm{h}$ exchanged with the hot bath}
\label{sec_expt}

The experimental challenge in implementing our proposed protocol lies in the accurate and precise measurement of the heat exchange $\Qh$, which is a difficult task under presently available experimental resources. As such, an alternative way to implement our proposed method would be to measure $\Qh$ indirectly through a state measurement of the known TLS $\mathcal{K}$ on either end of the thermalization stroke $\lt$; which can be done through various standard techniques, including motional sideband spectroscopy~\cite{safavi12} and photoluminescence~\cite{hopper18, tran2019anti}. In other words, we measure the excited state population of $\mathcal{K}$ before the stroke, which we denote as $n_\mathrm{k}'$ . The excited state population after thermalization  ($n_\mathrm{k}$) is directly inferred  from the temperature $T_\mathrm{h}$ and $\omegak$ using Eq.~\eqref{eq_denTLS}. $\Qh$ is then easily calculated from the measured values of $n_\mathrm{k}$ and $n_\mathrm{k}'$ (see Appendix~\ref{app_theta}). The error in $\Qh$ therefore solely arises from the measurement of the excited state population of the known TLS $\mathcal{K}$. To estimate the minimum error bound, we calculate the QFI of the state of $\mathcal{K}$ prior to thermalization with the excited state population $n_\mathrm{k}'$ as the parameter to be estimated (see Appendix~\ref{app_QFI}),   
\begin{equation}
\mathcal{F}_\mathrm{I}(n_k')\approx\frac{1}{n_\mathrm{un}(1-n_\mathrm{un})}=4\cosh^2(\omegaun/\Tc).
\end{equation}
The Cramer-Rao bound hence provides the lower bound on the error in $\Qh$ as~\cite{caves94, paris09quantum}
\begin{equation}\label{eq_alt_error}
\Delta \Qh\geq(1/2)\sech(\omegaun/\Tc),
\end{equation}
which is finite ($\sim 0.5$) even for $\omegaun/\Tc\to 0$. Further, multiple measurements reduce this lower bound as $1/\sqrt{\mathcal{M}_Q}$, where $\mathcal{M}_Q$ is the number of measurements performed.

\section{Discussion}
\label{sec_dis}

We propose a protocol for measuring a weak magnetic field by using a two-stroke thermal machine (TTM) magnetometer, modelled by a two-level system (TLS) whose energy spectrum depends on the unknown field coupled to a second TLS with known and tunable energy eigenvalues, and driving the combined system in a closed cycle between two thermal baths. We specifically studied two different types of interactions between the TLSs, one that leads to a swapping of the TLSs populations without any entanglement generation (swap TTM, Sec.~\ref{sec_swap}) and a second that generates an entangled state after the unitary stroke (mix TTM, Sec.~\ref{sec_mix}). 
However, as shown in Appendix.~\ref{app_heat}, the entanglement produced in the mix TTM does not affect the effectiveness of our protocol.

\par

Our method uses the presence of a transition point at which the thermal machine switches operation mode from a quantum heat engine to a quantum refrigerator. The transition is detected 
through a sign  reversal in the heat $\Qh$ extracted from the hot bath which makes it possible to detect the transition point more precisely than absolute value measurements. Our main result is that, 
under reasonable conditions, our protocol is capable of reducing the relative error in measuring the unknown energy gap $2\omegaun$ by a factor of $\Tc/\Th$. Through explicit calculation of the possible 
sources of error, we further showed that the protocol is most effective when for a rough estimate of $\omegaun$ the the temperature $\Tc$ of the cold bath is chosen such that $\omegaun/\Tc\approx 1.2$. 
This optimum ratio is a requirement in both considered TTMs and stems from the minimization of the theoretical lower bound of the relative error set by the quantum Cramer-Rao inequality. Furthermore, we 
emphasize that a major advantage of the proposed protocol is the possibility of accurately estimating a weak magnetic field acting on the qubit $\mathcal{U}$ by probing the auxiliary
qubit $\mathcal{K}$. This scheme of indirectly measuring a weak magnetic field can be highly beneficial in scenarios where existing constraints may prevent us from measuring the magnetic field directly, without 
affecting the value of the field.

\par

A possible application of our protocol is nanometre-scale high-resolution magnetometry where the two TLSs could be implemented by two non-interacting ions
or atoms, possibly of two different species, trapped in a single well of an optical lattice. Alternatively, the TLSs can be modelled by superconducting qubits
\cite{vion02manipulating, wallraff04strong, koch07charge}. These systems can be extremely well controlled experimentally~\cite{grimm2000optical,schindler2013quantum} and 
have been used in recent experimental implementations of quantum heat engines~\cite{rossnagel16a,vonlindenfels2019spin}. An alternative route may be spin systems based on 
nuclear-magnetic resonances or nitrogen-vacancy setups as in the QHE experiments from Refs.~\cite{peterson2018experimental,klatzow19experimental}. For future work we envisage
to adapt our protocol to autonomous quantum heat engines that do not require external control. A prime example may be the three-level maser coupled to a single cavity-field 
mode~\cite{scovil1959three}. The engine transition then corresponds to the lasing threshold and instead of measuring energy currents one could base the magnetometry scheme on the
measurement of the photon bunching parameter~\cite{wallsbook,niedenzu2019concepts}. The change in work capacity or ergotropy \cite{pusz78, lenard78, niedenzu18quantum} of a quantum system (quantum battery) \cite{binder15quantacell, campaioli17enhancing}, for example a quantum harmonic oscillator, coupled to the thermal machine, is then an indication for work output of the machine \cite{klimovsky15work, klimovsky14heat, niedenzu2019concepts}.  One could then consider estimating the weak magnetic field through measurement of the zero of the work output, by measuring the change in ergotropy of  the quantum battery. Such autonomous magnetometers based on microscopic thermal machines might enable us to perform high-precision magnetometry 
in biological systems at the sub-cellular level \cite{kucsko13nanometre}. As an estimate, a cold bath with $\Tc \sim$ mK, for example realized using laser cooling \cite{metcalf07laser}, a hot
bath at room temperature with $\Th \sim 10^{2}$ K and a
strong known magnetic field $\sim 100$ Gs, which for example can be realized using a small bar magnet, would allow us to estimate weak magnetic fields of the order of $10^{-3}$ Gs 
[see Eq. \eqref{eq_wun}].

\section*{Acknowledgments} S.\,B., U.\,B. and A.\,D. acknowledge Souvik Bandyopadhyay and Somnath Maity for comments. S.\,B. acknowledges CSIR, India for financial support. W.\,N.\ acknowledges support from an ESQ fellowship of the Austrian Academy of Sciences (\"OAW). AD acknowledges SPARC program, MHRD, India for financial support.

\appendix

  \section{Unimodality of the unknown energy gap}\label{app_unimodal}

The error analysis in Eq.~\eqref{eq_Delta_wun} requires $\omegaun$ to be unimodal. From Eq.~\eqref{eq_wun} and the condition
  \begin{equation}\label{eq_uni}
    \omegaun=\omegak^*\frac{\Tc}{\Th}=y
  \end{equation}
  for some fixed $y>0$, the probability distribution of $\omegaun$ at point $y$ is
  \begin{equation}\label{eq_app_P}
    P_{\omegaun}(y)=\int_0^\infty\mathrm{d} \Tc \int_0^\infty\mathrm{d} \Th P_\mathrm{k}\left(\frac{\Th y}{\Tc}\right)P_\mathrm{h}(\Th)P_\mathrm{c}(\Tc),
  \end{equation}
  where $P_\mathrm{k}(\omegak^*)$ denotes the distribution of $\omegak^*$, $P_\mathrm{h}(\Th)$ the distribution of $\Th$ and $P_\mathrm{c}(\Tc)$ the distribution of $\Tc$, respectively. 
  The rationale behind Eq.~\eqref{eq_app_P} is that these three quantities can be measured independently and are therefore stochastically independent. Further, 
  the probability distributions themselves can be considered to be Gaussian in the limit of large number of repeated and uncorrelated measurements performed for each of the quantity.

\par

  We have checked numerically for a variety of Gaussian distributions of $\omegak^*$, $\Tc$ and $\Th$ that the resulting distribution~\eqref{eq_app_P} is indeed unimodal, 
  i.e., exhibits a single peak in $y$.

  \section{Master equation approach}\label{app_heat}

 Within the Born, Markov and secular approximations~\cite{breuerbook}, the dynamics of the joint density operator $\rho_\mathrm{tot}$ during the unitary stroke for the mix TTM is governed by the 
 Lindblad master equation $\dot{\rho}_\mathrm{tot}(t)=L_T(\rt(t))$ where the propagator $L_T$ defined as
\begin{equation}\label{eq_propagator}
L_T(\rt(t))=-i[H_0,\rt(t)]+D_\mathrm{h}(\rt(t)) + D_\mathrm{c}(\rt(t)),
\end{equation}
where $H_0=\Hk+\Hun$ and the dissipators 
\begin{multline}\label{eq_dissi}
D_\mathrm{h(c)}(\rt(t))\\=l_\mathrm{k(un)}^-\rt(t)l_\mathrm{k(un)}^+-\frac{1}{2}\{l_\mathrm{k(un)}^+l_\mathrm{k(un)}^-,\rt(t)\}\\+e^{-\frac{2\omega_\mathrm{k(un)}}{T_\mathrm{h(c)}}}\left(l_\mathrm{k(un)}^+\rt(t)l_\mathrm{k(un)}^--\frac{1}{2}\{l_\mathrm{k(un)}^-l_\mathrm{k(un)}^+,\rt(t)\}\right)
\end{multline}
describe the dissipative interaction of $\mathcal{K}$ ($\mathcal{U}$) with the thermal bath $\Th$ ($\Tc$). Here we have defined the jump operators
\begin{subequations}
	\begin{align}
	l_\mathrm{k}^{\pm}&=\sqrt{\gamma_\mathrm{k}}(\sigma_{k}^{\pm}\otimes \mathcal{I}_\mathrm{un})\\
	l_\mathrm{un}^{\pm}&=\sqrt{\gamma_\mathrm{un}}(\mathcal{I}_\mathrm{k} \otimes \sigma_\mathrm{un}^{\pm}),
	\end{align}
\end{subequations}
where $\sigma_{k}^{\pm}=\left(\sigma_\mathrm{k}^x\pm i\sigma_\mathrm{k}^y\right)$ and $\sigma_\mathrm{un}^{\pm}=\left(\sigma_\mathrm{un}^x\pm i\sigma_\mathrm{un}^y\right)$; $\gamma_\mathrm{k}$ and $\gamma_\mathrm{un}$ are the associated decay rates. The heat~\cite{alicki1979quantum}
\begin{equation}\label{eq_Q}
Q = \int_{\tau_\mathrm{U}}^{\tau_\mathrm{U}+\tau_\mathrm{T}}{\Tr}[\dot{\rho}_\mathrm{tot}(t)(\Hk+\Hun)]\mathrm{d}t
\end{equation}
exchanged with the baths during the stroke $\lt$ is decomposed into the individual contributions
\begin{equation}\label{eq_QE}
Q_\mathrm{h(c)}= \int_{\tau_\mathrm{U}}^{\tau_\mathrm{U}+\tau_\mathrm{T}}\Tr[D_\mathrm{h(c)}[\rt(t)](\Hk+\Hun)]\mathrm{d}t
\end{equation}
pertaining to the two baths $\Th$ and $\Tc$, respectively.

\par

However, the heat exchanges defined above evaluate to Eq.~\eqref{eq_mix_heat} since entanglement does not contribute to the heat exchanged between the combined system and the baths. 
To show this rigorously, let us recall that $H_\mathrm{I}(t)$ chosen for the mix TTM [Eq.~\eqref{eq_HI}] only causes a rotation of the projection of the combined system density matrix
$\rho_\mathrm{tot}$ on the subspace spanned by the states $\ket{\uparrow_\mathrm{k}}\ket{\downarrow_\mathrm{un}}$ and $\ket{\downarrow_\mathrm{k}}\ket{\uparrow_\mathrm{un}}$. As such, the general 
form of $\rho_\mathrm{tot}$ at any time is
\begin{equation}\label{eq_app_rho}
  \rho_\mathrm{tot}=\begin{pmatrix}
    a & 0 & 0 & 0\\
    0 & b & c & 0\\
    0 & c^* & d & 0\\
    0 & 0 & 0 & e
  \end{pmatrix},
\end{equation}
which can be rewritten
\begin{multline}
  \rho_\mathrm{tot}= \rho_\mathrm{k}\otimes\rho_\mathrm{un} - (bd-ae)\left(\sigma_z^\mathrm{k}\otimes\sigma_z^\mathrm{un}\right)\\ + c\left(\sigma_-^\mathrm{k}\otimes\sigma_+^\mathrm{un}\right) + c^*\left(\sigma_+^\mathrm{k}\otimes\sigma_-^\mathrm{un}\right),
\end{multline}
where 
\begin{equation}
  \rho_\mathrm{k(un)}=\Tr_\mathrm{un(k)}[\rho_\mathrm{tot}]
\end{equation}
The total energy contained in the TLSs is therefore
\begin{align}
  E&=\Tr[\rho_\mathrm{tot}H_0]\notag\\
   &=\Tr[(\rho_\mathrm{k}\otimes\rho_\mathrm{un})H_0] - (bd-ae)\Tr[\left(\sigma_z^\mathrm{k}\otimes\sigma_z^\mathrm{un}\right)H_0]\notag\\
   &\quad+c\Tr[\left(\sigma_-^\mathrm{k}\otimes\sigma_+^\mathrm{un}\right)H_0] + c^*\Tr[\left(\sigma_+^\mathrm{k}\otimes\sigma_-^\mathrm{un}\right)H_0].
\end{align}

We evaluate the constituents of the above equation as follows: Considering the first term, we have
\begin{align}
  \Tr[(\rho_\mathrm{k}\otimes&\rho_\mathrm{un})H_0]\nonumber\\& =\Tr[(\rho_\mathrm{k}\otimes\rho_\mathrm{un})(\Hk\otimes\mathcal{I}_\mathrm{un}+\mathcal{I}_\mathrm{k}\otimes \Hun)]\nonumber\\
  &=\Tr[\rho_\mathrm{k}\Hk\otimes\rho_\mathrm{un}]+\Tr[\rho_\mathrm{k}\otimes\rho_\mathrm{un}\Hun]\nonumber \\
  &=\Tr[\rho_\mathrm{k}\Hk]\Tr[\rho_\mathrm{un}] + \Tr[\rho_\mathrm{k}]\Tr[\rho_\mathrm{un}\Hun]\nonumber\\
  &=\Tr[\rho_\mathrm{k}\Hk] + \Tr[\rho_\mathrm{un}\Hun].
\end{align}
Similarly, the second term evaluates to
\begin{align}
  &(bd-ae)\Tr[\left(\sigma_z^\mathrm{k}\otimes\sigma_z^\mathrm{un}\right)H_0]\nonumber\\&=(bd-ae)\left(\Tr[\sigma_z^\mathrm{k}\Hk]\Tr[\sigma_z^\mathrm{un}] + \Tr[\sigma_z^\mathrm{k}]\Tr[\sigma_z^\mathrm{un}\Hun]\right)\nonumber\\&=0
\end{align} 
since $\Tr[\sigma_z^\mathrm{k}]=\Tr[\sigma_z^\mathrm{un}]=0$. Similarly, it is straightforward to show that the third and fourth terms also vanish. Hence, Eq.~\eqref{eq_Q} evaluates to
\begin{align}
  Q &= \int_{\tau_\mathrm{U}}^{\tau_\mathrm{U}+\tau_\mathrm{T}}\Tr[\dot{\rho}_\mathrm{tot}(t)(\Hk+\Hun)]\mathrm{d}t \nonumber\\
    &=\int_{\tau_\mathrm{U}}^{\tau_\mathrm{U}+\tau_\mathrm{T}}\Tr[\dot{\rho}_\mathrm{k}(t)\Hk]\mathrm{d}t + \int_{\tau_\mathrm{U}}^{\tau_\mathrm{U}+\tau_{T}}\Tr[\dot{\rho}_\mathrm{un}(t)\Hun]\mathrm{d}t\nonumber\\
    &=\left(\Tr[\rho_\mathrm{k}(\tau_\mathrm{U}+\tau_\mathrm{T})H_\mathrm{k}] - \Tr[\rho_\mathrm{k}(\tau_\mathrm{U})H_\mathrm{k}]\right)\nonumber\\ & \quad+\left(\Tr[\rho_\mathrm{un}(\tau_\mathrm{U}+\tau_\mathrm{T})\Hun] - \Tr[\rho_\mathrm{un}(\tau_\mathrm{U})\Hun]\right),
\end{align}
which is Eq.~\eqref{eq_mix_heat}.

\section{Heat exchanges and work done for an arbitrary mixing angle}\label{app_theta}
As discussed in Appendix~\ref{app_heat}, the stroke $\lu$ only rotates the projection of the density matrix of the combined system in the subspace spanned by the states $\ket{\uparrow_\mathrm{k}}\ket{\downarrow_\mathrm{un}}$ and $\ket{\downarrow_\mathrm{k}}\ket{\uparrow_\mathrm{un}}$. For an arbitrary rotation of angle $\theta$, with the initial density matrix given by Eq.~\eqref{eq_rho_tot}, the evolved density matrix assumes the form 
\begin{widetext}
\begin{equation}
\rho_\mathrm{tot}(\theta)=\begin{pmatrix}
n_\mathrm{k}n_\mathrm{un} & 0 & 0 & 0\\
0 &  n_\mathrm{k}\cos^2\theta+n_\mathrm{un}\sin^2\theta -n_\mathrm{k}n_\mathrm{un}& (n_\mathrm{k}-n_\mathrm{un})\sin \theta \cos\theta& 0\\
0 & (n_\mathrm{k}-n_\mathrm{un})\sin \theta \cos\theta &n_\mathrm{k}\sin^2\theta+n_\mathrm{un}\cos^2\theta -n_\mathrm{k}n_\mathrm{un}& 0\\
0 & 0 & 0 & (1-n_\mathrm{k})(1-n_\mathrm{un})
\end{pmatrix}.
\end{equation}
\end{widetext}
The reduced density matrices of the the TLSs hence remain diagonal and is given as
\begin{subequations}\label{eq_app_denTLS}
	\begin{gather}
	\rho_\mathrm{k}(\tau_{U})=\begin{pmatrix}   
	n_\mathrm{k}' & 0\\
	0 & 1-n_\mathrm{k}'
	\end{pmatrix},
	\rho_\mathrm{un}(\tau_{U})=
	\begin{pmatrix}   
	n_\mathrm{un}' & 0\\
	0 & 1-n_\mathrm{un}'
	\end{pmatrix},
	\end{gather}
\end{subequations} 
where,
\begin{subequations}
\begin{align}
n_\mathrm{k}'=n_\mathrm{k}\cos^2\theta+n_\mathrm{un}\sin^2\theta,
\end{align}	
\begin{align}
n_\mathrm{un}'=n_\mathrm{k}\sin^2\theta+n_\mathrm{un}\cos^2\theta.
\end{align}
\end{subequations}

Using Eqs.~\eqref{eq_work}, \eqref{eq_mix_heat} and \eqref{eq_rku_mix}, one can therefore obtain
\begin{subequations} 
\begin{align}
\Qh&~=2\omegak(n_\mathrm{k}-n_\mathrm{k}')=2\omegak(n_\mathrm{k}-n_\mathrm{un})\sin^2{\theta}\\
\Qc&~=2\omegaun(n_\mathrm{un}-n_\mathrm{un}')=2\omegaun(n_\mathrm{un}-n_\mathrm{k})\sin^2{\theta}\\
W&=-2(\omegak-\omegaun)(n_\mathrm{k}-n_\mathrm{un})\sin^2{\theta}.
\end{align}
\end{subequations}

\section{Thermalization time for stroke $\lt$}\label{app_therm}

As discussed in the main text, the timescale on which the combined system of the two TLSs thermalizes is determined by
$\tau_\mathrm{th}=\min\{|\realp[\lambda_i]|\neq 0\}^{-1}\propto\gamma^{-1}$, where $\lambda_i$ are the eigenvalues of the propagator for the stroke $\lt$ defined in Eq.~\eqref{eq_propagator}, which 
evaluates to 
\begin{equation}
\tau_\mathrm{th}=\min\{\gamma_\mathrm{k}(1+e^{-2\w_\mathrm{k}/\Th}),\gamma_\mathrm{un}(1+e^{-2\w_\mathrm{un}/\Tc})\}^{-1}.
\end{equation}
For the purpose of simplicity we assume equal decay rates for both TLS, $\gamma_\mathrm{k}=\gamma_\mathrm{un}=\gamma$. Also, since the TTM will be operated in the vicinity of the transition 
point we have $\w_\mathrm{k}/\Th\approx\w_\mathrm{un}/\Tc$ and, consequently, $\tau_\mathrm{th}\approx\gamma^{-1}(1+e^{-2\w_\mathrm{un}/\Tc})^{-1}$.

\section{Calculation for quantum Fisher information}\label{app_QFI}

For a system initialized in the state $\rho=\sum_{n}p_n\ket{\psi_n}\bra{\psi_n}$, the QFI (denoted by $\mathcal{F}_\mathrm{I}$) as defined in Eq.~\eqref{eq_qfi} can be expanded into the following form~\cite{paris09quantum}
\begin{equation}
\mathcal{F}_\mathrm{I}(\lambda)=\sum_{i=1}\frac{(\partial_{\lambda}p_i)^2}{p_i} + 2\sum_{i\neq j}^Nc_{ij}\left|\la\psi_j|\partial_{\lambda}\psi_i\ra\right|^2,
\end{equation}
where $\lambda$ is the parameter to be estimated and
\begin{equation}
c_{ij}=2p_i\frac{p_i-p_j}{p_i+p_j}.
\end{equation}

Using the above definition, we calculate the QFI for the following cases:

\paragraph{For the TLS $\mathcal{U}$ in thermal equilibrium with bath $\Tc$ and $\omegaun$ as the parameter to be estimated:} 
In this case, the density matrix $\rho_\mathrm{un}$ is diagonal in the energy eigenbasis, $\rho_\mathrm{un}=n_\mathrm{un}\ket{\uparrow_\mathrm{un}}\bra{\uparrow_\mathrm{un}} + (1-n_\mathrm{un})\ket{\downarrow_\mathrm{un}}\bra{\downarrow_\mathrm{un}}$. Note that the eigenvectors $\ket{\uparrow_\mathrm{un}}$ and $\ket{\downarrow_\mathrm{un}}$ are independent of the magnitude of $\omegaun$ and hence the QFI reduces to
\begin{align}
  \mathcal{F}_\mathrm{I}(\omegaun)&=\frac{1}{n_\mathrm{un}}\left(\frac{\partial n_\mathrm{un}}{\partial \omegaun}\right)^2 + \frac{1}{1-n_\mathrm{un}}\left(\frac{\partial (1-n_\mathrm{un})}{\partial \omegaun}\right)^2\notag\\
                                  &=\frac{1}{n_\mathrm{un}(1-n_\mathrm{un})}\left(\frac{\partial n_\mathrm{un}}{\partial \omegaun}\right)^2
\end{align}
Substituting
\begin{equation}
  n_\mathrm{un}=\frac{e^{-\omegaun/\Tc}}{e^{-\omegaun/\Tc}+e^{\omegaun/\Tc}},
\end{equation}
we obtain
\begin{equation}
  \mathcal{F}_\mathrm{I}(\omegaun)=\frac{1}{T^2}\sech^2{\left(\frac{\omegaun}{T}\right)}. \label{eq_App3}
\end{equation}
Using the $\mathcal{F}_\mathrm{I}$ calculated above, the quantum Cramer-Rao bound is obtained in Eq.~\eqref{eq_crbound}.

\paragraph{For the TLS $\mathcal{K}$ before the thermalization stroke and $n_\mathrm{k}'$ as the parameter to be estimated:} 
As discussed in Appendix~\ref{app_theta}, the density matrix of $\mathcal{K}$ after the thermalization stroke assumes the form
\begin{equation}
\rho_k(\tau_U)=\begin{pmatrix}
n_k' & 0\\
0 & 1-n_k'
\end{pmatrix}.
\end{equation}
The QFI is therefore obtained as
\begin{align}
\mathcal{F}_\mathrm{I}(n_k')&=\frac{1}{n_k'}\left(\frac{\partial n_k'}{\partial n_k'}\right)^2 + \frac{1}{1-n_k'}\left(\frac{\partial (1-n_k')}{\partial n_k'}\right)^2\notag\\
&=\frac{1}{n_k'(1-n_k')}.
\end{align}
As our measurements are performed close to the transition point, we have
\begin{equation}
\lim_{\omegak\to\omegak^*}n_k'=n_k (\omegak^*)=n_\mathrm{un}.
\end{equation}
Consequently, the QFI reduces to
\begin{equation}
\mathcal{F}_\mathrm{I}(n_k')=\frac{1}{n_\mathrm{un}(1-n_\mathrm{un})}=4\cosh^2(\omegaun/\Tc),
\end{equation}
from which the Cramer-Rao bound in Eq.~\eqref{eq_alt_error} can be obtained.


\begin{thebibliography}{86}%
\makeatletter
\providecommand \@ifxundefined [1]{%
 \@ifx{#1\undefined}
}%
\providecommand \@ifnum [1]{%
 \ifnum #1\expandafter \@firstoftwo
 \else \expandafter \@secondoftwo
 \fi
}%
\providecommand \@ifx [1]{%
 \ifx #1\expandafter \@firstoftwo
 \else \expandafter \@secondoftwo
 \fi
}%
\providecommand \natexlab [1]{#1}%
\providecommand \enquote  [1]{``#1''}%
\providecommand \bibnamefont  [1]{#1}%
\providecommand \bibfnamefont [1]{#1}%
\providecommand \citenamefont [1]{#1}%
\providecommand \href@noop [0]{\@secondoftwo}%
\providecommand \href [0]{\begingroup \@sanitize@url \@href}%
\providecommand \@href[1]{\@@startlink{#1}\@@href}%
\providecommand \@@href[1]{\endgroup#1\@@endlink}%
\providecommand \@sanitize@url [0]{\catcode `\\12\catcode `\$12\catcode
  `\&12\catcode `\#12\catcode `\^12\catcode `\_12\catcode `\%12\relax}%
\providecommand \@@startlink[1]{}%
\providecommand \@@endlink[0]{}%
\providecommand \url  [0]{\begingroup\@sanitize@url \@url }%
\providecommand \@url [1]{\endgroup\@href {#1}{\urlprefix }}%
\providecommand \urlprefix  [0]{URL }%
\providecommand \Eprint [0]{\href }%
\providecommand \doibase [0]{http://dx.doi.org/}%
\providecommand \selectlanguage [0]{\@gobble}%
\providecommand \bibinfo  [0]{\@secondoftwo}%
\providecommand \bibfield  [0]{\@secondoftwo}%
\providecommand \translation [1]{[#1]}%
\providecommand \BibitemOpen [0]{}%
\providecommand \bibitemStop [0]{}%
\providecommand \bibitemNoStop [0]{.\EOS\space}%
\providecommand \EOS [0]{\spacefactor3000\relax}%
\providecommand \BibitemShut  [1]{\csname bibitem#1\endcsname}%
\let\auto@bib@innerbib\@empty
\bibitem [{\citenamefont {Nickerson}\ \emph {et~al.}(2014)\citenamefont
  {Nickerson}, \citenamefont {Fitzsimons},\ and\ \citenamefont
  {Benjamin}}]{Nickerson14freely}%
  \BibitemOpen
  \bibfield  {author} {\bibinfo {author} {\bibfnamefont {N.~H.}\ \bibnamefont
  {Nickerson}}, \bibinfo {author} {\bibfnamefont {J.~F.}\ \bibnamefont
  {Fitzsimons}}, \ and\ \bibinfo {author} {\bibfnamefont {S.~C.}\ \bibnamefont
  {Benjamin}},\ }\href {\doibase 10.1103/PhysRevX.4.041041} {\bibfield
  {journal} {\bibinfo  {journal} {Phys. Rev. X}\ }\textbf {\bibinfo {volume}
  {4}},\ \bibinfo {pages} {041041} (\bibinfo {year} {2014})}\BibitemShut
  {NoStop}%
\bibitem [{\citenamefont {Kurizki}\ \emph {et~al.}(2015)\citenamefont
  {Kurizki}, \citenamefont {Bertet}, \citenamefont {Kubo}, \citenamefont
  {M{\o}lmer}, \citenamefont {Petrosyan}, \citenamefont {Rabl},\ and\
  \citenamefont {Schmiedmayer}}]{kurizki15quantum}%
  \BibitemOpen
  \bibfield  {author} {\bibinfo {author} {\bibfnamefont {G.}~\bibnamefont
  {Kurizki}}, \bibinfo {author} {\bibfnamefont {P.}~\bibnamefont {Bertet}},
  \bibinfo {author} {\bibfnamefont {Y.}~\bibnamefont {Kubo}}, \bibinfo {author}
  {\bibfnamefont {K.}~\bibnamefont {M{\o}lmer}}, \bibinfo {author}
  {\bibfnamefont {D.}~\bibnamefont {Petrosyan}}, \bibinfo {author}
  {\bibfnamefont {P.}~\bibnamefont {Rabl}}, \ and\ \bibinfo {author}
  {\bibfnamefont {J.}~\bibnamefont {Schmiedmayer}},\ }\href {\doibase
  10.1073/pnas.1419326112} {\bibfield  {journal} {\bibinfo  {journal} {Proc.
  Natl. Acad. Sci. USA}\ }\textbf {\bibinfo {volume} {112}},\ \bibinfo {pages}
  {3866} (\bibinfo {year} {2015})}\BibitemShut {NoStop}%
\bibitem [{\citenamefont {Ro{\ss}nagel}\ \emph {et~al.}(2016)\citenamefont
  {Ro{\ss}nagel}, \citenamefont {Dawkins}, \citenamefont {Tolazzi},
  \citenamefont {Abah}, \citenamefont {Lutz}, \citenamefont {Schmidt-Kaler},\
  and\ \citenamefont {Singer}}]{rossnagel16a}%
  \BibitemOpen
  \bibfield  {author} {\bibinfo {author} {\bibfnamefont {J.}~\bibnamefont
  {Ro{\ss}nagel}}, \bibinfo {author} {\bibfnamefont {S.~T.}\ \bibnamefont
  {Dawkins}}, \bibinfo {author} {\bibfnamefont {K.~N.}\ \bibnamefont
  {Tolazzi}}, \bibinfo {author} {\bibfnamefont {O.}~\bibnamefont {Abah}},
  \bibinfo {author} {\bibfnamefont {E.}~\bibnamefont {Lutz}}, \bibinfo {author}
  {\bibfnamefont {F.}~\bibnamefont {Schmidt-Kaler}}, \ and\ \bibinfo {author}
  {\bibfnamefont {K.}~\bibnamefont {Singer}},\ }\href {\doibase
  10.1126/science.aad6320} {\bibfield  {journal} {\bibinfo  {journal}
  {Science}\ }\textbf {\bibinfo {volume} {352}},\ \bibinfo {pages} {325}
  (\bibinfo {year} {2016})}\BibitemShut {NoStop}%
\bibitem [{\citenamefont {Bernien}\ \emph {et~al.}(2017)\citenamefont
  {Bernien}, \citenamefont {Schwartz}, \citenamefont {Keesling}, \citenamefont
  {Levine}, \citenamefont {Omran}, \citenamefont {Pichler}, \citenamefont
  {Choi}, \citenamefont {Zibrov}, \citenamefont {Endres}, \citenamefont
  {Greiner}, \citenamefont {Vuleti\'{c}},\ and\ \citenamefont
  {Lukin}}]{bernien17probing}%
  \BibitemOpen
  \bibfield  {author} {\bibinfo {author} {\bibfnamefont {H.}~\bibnamefont
  {Bernien}}, \bibinfo {author} {\bibfnamefont {S.}~\bibnamefont {Schwartz}},
  \bibinfo {author} {\bibfnamefont {A.}~\bibnamefont {Keesling}}, \bibinfo
  {author} {\bibfnamefont {H.}~\bibnamefont {Levine}}, \bibinfo {author}
  {\bibfnamefont {A.}~\bibnamefont {Omran}}, \bibinfo {author} {\bibfnamefont
  {H.}~\bibnamefont {Pichler}}, \bibinfo {author} {\bibfnamefont
  {S.}~\bibnamefont {Choi}}, \bibinfo {author} {\bibfnamefont {A.~S.}\
  \bibnamefont {Zibrov}}, \bibinfo {author} {\bibfnamefont {M.}~\bibnamefont
  {Endres}}, \bibinfo {author} {\bibfnamefont {M.}~\bibnamefont {Greiner}},
  \bibinfo {author} {\bibfnamefont {V.}~\bibnamefont {Vuleti\'{c}}}, \ and\
  \bibinfo {author} {\bibfnamefont {M.~D.}\ \bibnamefont {Lukin}},\ }\href
  {\doibase 10.1038/nature24622} {\bibfield  {journal} {\bibinfo  {journal}
  {Nature}\ }\textbf {\bibinfo {volume} {551}},\ \bibinfo {pages} {579}
  (\bibinfo {year} {2017})}\BibitemShut {NoStop}%
\bibitem [{\citenamefont {Paris}(2009)}]{paris09quantum}%
  \BibitemOpen
  \bibfield  {author} {\bibinfo {author} {\bibfnamefont {M.~G.~A.}\
  \bibnamefont {Paris}},\ }\href {\doibase 10.1142/S0219749909004839}
  {\bibfield  {journal} {\bibinfo  {journal} {Int. J. Quantum Inf.}\ }\textbf
  {\bibinfo {volume} {07}},\ \bibinfo {pages} {125} (\bibinfo {year}
  {2009})}\BibitemShut {NoStop}%
\bibitem [{\citenamefont {Correa}\ \emph {et~al.}(2015)\citenamefont {Correa},
  \citenamefont {Mehboudi}, \citenamefont {Adesso},\ and\ \citenamefont
  {Sanpera}}]{correa15individual}%
  \BibitemOpen
  \bibfield  {author} {\bibinfo {author} {\bibfnamefont {L.~A.}\ \bibnamefont
  {Correa}}, \bibinfo {author} {\bibfnamefont {M.}~\bibnamefont {Mehboudi}},
  \bibinfo {author} {\bibfnamefont {G.}~\bibnamefont {Adesso}}, \ and\ \bibinfo
  {author} {\bibfnamefont {A.}~\bibnamefont {Sanpera}},\ }\href {\doibase
  10.1103/PhysRevLett.114.220405} {\bibfield  {journal} {\bibinfo  {journal}
  {Phys. Rev. Lett.}\ }\textbf {\bibinfo {volume} {114}},\ \bibinfo {pages}
  {220405} (\bibinfo {year} {2015})}\BibitemShut {NoStop}%
\bibitem [{\citenamefont {Zwick}\ \emph {et~al.}(2016)\citenamefont {Zwick},
  \citenamefont {\'Alvarez},\ and\ \citenamefont
  {Kurizki}}]{zwick16criticality}%
  \BibitemOpen
  \bibfield  {author} {\bibinfo {author} {\bibfnamefont {A.}~\bibnamefont
  {Zwick}}, \bibinfo {author} {\bibfnamefont {G.~A.}\ \bibnamefont
  {\'Alvarez}}, \ and\ \bibinfo {author} {\bibfnamefont {G.}~\bibnamefont
  {Kurizki}},\ }\href {\doibase 10.1103/PhysRevA.94.042122} {\bibfield
  {journal} {\bibinfo  {journal} {Phys. Rev. A}\ }\textbf {\bibinfo {volume}
  {94}},\ \bibinfo {pages} {042122} (\bibinfo {year} {2016})}\BibitemShut
  {NoStop}%
\bibitem [{\citenamefont {Zhou}\ \emph {et~al.}(2018)\citenamefont {Zhou},
  \citenamefont {Zhang}, \citenamefont {Preskill},\ and\ \citenamefont
  {Jiang}}]{zhou18achieving}%
  \BibitemOpen
  \bibfield  {author} {\bibinfo {author} {\bibfnamefont {S.}~\bibnamefont
  {Zhou}}, \bibinfo {author} {\bibfnamefont {M.}~\bibnamefont {Zhang}},
  \bibinfo {author} {\bibfnamefont {J.}~\bibnamefont {Preskill}}, \ and\
  \bibinfo {author} {\bibfnamefont {L.}~\bibnamefont {Jiang}},\ }\href
  {\doibase 10.1038/s41467-017-02510-3} {\bibfield  {journal} {\bibinfo
  {journal} {Nat. Comm.}\ }\textbf {\bibinfo {volume} {9}},\ \bibinfo {pages}
  {78} (\bibinfo {year} {2018})}\BibitemShut {NoStop}%
\bibitem [{\citenamefont {Kacprowicz}\ \emph {et~al.}(2010)\citenamefont
  {Kacprowicz}, \citenamefont {Demkowicz-Dobrzanski}, \citenamefont
  {Wasilewski}, \citenamefont {Banaszek},\ and\ \citenamefont
  {Walmsley}}]{kacprowicz10experimental}%
  \BibitemOpen
  \bibfield  {author} {\bibinfo {author} {\bibfnamefont {M.}~\bibnamefont
  {Kacprowicz}}, \bibinfo {author} {\bibfnamefont {R.}~\bibnamefont
  {Demkowicz-Dobrzanski}}, \bibinfo {author} {\bibfnamefont {W.}~\bibnamefont
  {Wasilewski}}, \bibinfo {author} {\bibfnamefont {K.}~\bibnamefont
  {Banaszek}}, \ and\ \bibinfo {author} {\bibfnamefont {I.~A.}\ \bibnamefont
  {Walmsley}},\ }\href {\doibase 10.1038/nphoton.2010.39} {\bibfield  {journal}
  {\bibinfo  {journal} {Nat. Photonics}\ }\textbf {\bibinfo {volume} {4}},\
  \bibinfo {pages} {357} (\bibinfo {year} {2010})}\BibitemShut {NoStop}%
\bibitem [{\citenamefont {Kucsko}\ \emph
  {et~al.}(2013{\natexlab{a}})\citenamefont {Kucsko}, \citenamefont {Maurer},
  \citenamefont {Yao}, \citenamefont {Kubo}, \citenamefont {Noh}, \citenamefont
  {Lo}, \citenamefont {Park},\ and\ \citenamefont {Lukin}}]{kucksko13}%
  \BibitemOpen
  \bibfield  {author} {\bibinfo {author} {\bibfnamefont {G.}~\bibnamefont
  {Kucsko}}, \bibinfo {author} {\bibfnamefont {P.~C.}\ \bibnamefont {Maurer}},
  \bibinfo {author} {\bibfnamefont {N.~Y.}\ \bibnamefont {Yao}}, \bibinfo
  {author} {\bibfnamefont {M.}~\bibnamefont {Kubo}}, \bibinfo {author}
  {\bibfnamefont {H.~J.}\ \bibnamefont {Noh}}, \bibinfo {author} {\bibfnamefont
  {P.~K.}\ \bibnamefont {Lo}}, \bibinfo {author} {\bibfnamefont
  {H.}~\bibnamefont {Park}}, \ and\ \bibinfo {author} {\bibfnamefont {M.~D.}\
  \bibnamefont {Lukin}},\ }\href {\doibase 10.1038/nature12373} {\bibfield
  {journal} {\bibinfo  {journal} {Nature}\ }\textbf {\bibinfo {volume} {500}},\
  \bibinfo {pages} {54} (\bibinfo {year} {2013}{\natexlab{a}})}\BibitemShut
  {NoStop}%
\bibitem [{\citenamefont {Toyli}\ \emph {et~al.}(2013)\citenamefont {Toyli},
  \citenamefont {de~las Casas}, \citenamefont {Christle}, \citenamefont
  {Dobrovitski},\ and\ \citenamefont {Awschalom}}]{toyli13fluorescence}%
  \BibitemOpen
  \bibfield  {author} {\bibinfo {author} {\bibfnamefont {D.~M.}\ \bibnamefont
  {Toyli}}, \bibinfo {author} {\bibfnamefont {C.~F.}\ \bibnamefont {de~las
  Casas}}, \bibinfo {author} {\bibfnamefont {D.~J.}\ \bibnamefont {Christle}},
  \bibinfo {author} {\bibfnamefont {V.~V.}\ \bibnamefont {Dobrovitski}}, \ and\
  \bibinfo {author} {\bibfnamefont {D.~D.}\ \bibnamefont {Awschalom}},\ }\href
  {\doibase 10.1073/pnas.1306825110} {\bibfield  {journal} {\bibinfo  {journal}
  {Proc. Natl. Acad. Sci. USA}\ }\textbf {\bibinfo {volume} {110}},\ \bibinfo
  {pages} {8417} (\bibinfo {year} {2013})}\BibitemShut {NoStop}%
\bibitem [{\citenamefont {Giovannetti}\ \emph {et~al.}(2011)\citenamefont
  {Giovannetti}, \citenamefont {Lloyd},\ and\ \citenamefont
  {Maccone}}]{giovannetti11}%
  \BibitemOpen
  \bibfield  {author} {\bibinfo {author} {\bibfnamefont {V.}~\bibnamefont
  {Giovannetti}}, \bibinfo {author} {\bibfnamefont {S.}~\bibnamefont {Lloyd}},
  \ and\ \bibinfo {author} {\bibfnamefont {L.}~\bibnamefont {Maccone}},\ }\href
  {\doibase 10.1038/nphoton.2011.35} {\bibfield  {journal} {\bibinfo  {journal}
  {Nature Photonics}\ }\textbf {\bibinfo {volume} {5}},\ \bibinfo {pages} {222}
  (\bibinfo {year} {2011})}\BibitemShut {NoStop}%
\bibitem [{\citenamefont {Degen}\ \emph {et~al.}(2017)\citenamefont {Degen},
  \citenamefont {Reinhard},\ and\ \citenamefont {Cappellaro}}]{degen17quantum}%
  \BibitemOpen
  \bibfield  {author} {\bibinfo {author} {\bibfnamefont {C.~L.}\ \bibnamefont
  {Degen}}, \bibinfo {author} {\bibfnamefont {F.}~\bibnamefont {Reinhard}}, \
  and\ \bibinfo {author} {\bibfnamefont {P.}~\bibnamefont {Cappellaro}},\
  }\href {\doibase 10.1103/RevModPhys.89.035002} {\bibfield  {journal}
  {\bibinfo  {journal} {Rev. Mod. Phys.}\ }\textbf {\bibinfo {volume} {89}},\
  \bibinfo {pages} {035002} (\bibinfo {year} {2017})}\BibitemShut {NoStop}%
\bibitem [{\citenamefont {Kurizki}\ \emph {et~al.}(2017)\citenamefont
  {Kurizki}, \citenamefont {Alvarez},\ and\ \citenamefont
  {Zwick}}]{zwick17quantum}%
  \BibitemOpen
  \bibfield  {author} {\bibinfo {author} {\bibfnamefont {G.}~\bibnamefont
  {Kurizki}}, \bibinfo {author} {\bibfnamefont {G.~A.}\ \bibnamefont
  {Alvarez}}, \ and\ \bibinfo {author} {\bibfnamefont {A.}~\bibnamefont
  {Zwick}},\ }\href {\doibase 10.3390/technologies5010001} {\bibfield
  {journal} {\bibinfo  {journal} {Technologies}\ }\textbf {\bibinfo {volume}
  {5}},\ \bibinfo {pages} {1} (\bibinfo {year} {2017})}\BibitemShut {NoStop}%
\bibitem [{\citenamefont {Pezz\`e}\ \emph {et~al.}(2018)\citenamefont
  {Pezz\`e}, \citenamefont {Smerzi}, \citenamefont {Oberthaler}, \citenamefont
  {Schmied},\ and\ \citenamefont {Treutlein}}]{pezze18quantum}%
  \BibitemOpen
  \bibfield  {author} {\bibinfo {author} {\bibfnamefont {L.}~\bibnamefont
  {Pezz\`e}}, \bibinfo {author} {\bibfnamefont {A.}~\bibnamefont {Smerzi}},
  \bibinfo {author} {\bibfnamefont {M.~K.}\ \bibnamefont {Oberthaler}},
  \bibinfo {author} {\bibfnamefont {R.}~\bibnamefont {Schmied}}, \ and\
  \bibinfo {author} {\bibfnamefont {P.}~\bibnamefont {Treutlein}},\ }\href
  {\doibase 10.1103/RevModPhys.90.035005} {\bibfield  {journal} {\bibinfo
  {journal} {Rev. Mod. Phys.}\ }\textbf {\bibinfo {volume} {90}},\ \bibinfo
  {pages} {035005} (\bibinfo {year} {2018})}\BibitemShut {NoStop}%
\bibitem [{\citenamefont {Wineland}\ and\ \citenamefont
  {Leibfried}(2011)}]{wineland11quantum}%
  \BibitemOpen
  \bibfield  {author} {\bibinfo {author} {\bibfnamefont {D.}~\bibnamefont
  {Wineland}}\ and\ \bibinfo {author} {\bibfnamefont {D.}~\bibnamefont
  {Leibfried}},\ }\href {\doibase 10.1002/lapl.201010125} {\bibfield  {journal}
  {\bibinfo  {journal} {Laser Phys. Lett.}\ }\textbf {\bibinfo {volume} {8}},\
  \bibinfo {pages} {175} (\bibinfo {year} {2011})}\BibitemShut {NoStop}%
\bibitem [{\citenamefont {Li}\ and\ \citenamefont
  {Luo}(2013)}]{li13entanglement}%
  \BibitemOpen
  \bibfield  {author} {\bibinfo {author} {\bibfnamefont {N.}~\bibnamefont
  {Li}}\ and\ \bibinfo {author} {\bibfnamefont {S.}~\bibnamefont {Luo}},\
  }\href {\doibase 10.1103/PhysRevA.88.014301} {\bibfield  {journal} {\bibinfo
  {journal} {Phys. Rev. A}\ }\textbf {\bibinfo {volume} {88}},\ \bibinfo
  {pages} {014301} (\bibinfo {year} {2013})}\BibitemShut {NoStop}%
\bibitem [{\citenamefont {Strobel}\ \emph {et~al.}(2014)\citenamefont
  {Strobel}, \citenamefont {Muessel}, \citenamefont {Linnemann}, \citenamefont
  {Zibold}, \citenamefont {Hume}, \citenamefont {Pezz{\`e}}, \citenamefont
  {Smerzi},\ and\ \citenamefont {Oberthaler}}]{strobel14}%
  \BibitemOpen
  \bibfield  {author} {\bibinfo {author} {\bibfnamefont {H.}~\bibnamefont
  {Strobel}}, \bibinfo {author} {\bibfnamefont {W.}~\bibnamefont {Muessel}},
  \bibinfo {author} {\bibfnamefont {D.}~\bibnamefont {Linnemann}}, \bibinfo
  {author} {\bibfnamefont {T.}~\bibnamefont {Zibold}}, \bibinfo {author}
  {\bibfnamefont {D.~B.}\ \bibnamefont {Hume}}, \bibinfo {author}
  {\bibfnamefont {L.}~\bibnamefont {Pezz{\`e}}}, \bibinfo {author}
  {\bibfnamefont {A.}~\bibnamefont {Smerzi}}, \ and\ \bibinfo {author}
  {\bibfnamefont {M.~K.}\ \bibnamefont {Oberthaler}},\ }\href {\doibase
  10.1126/science.1250147} {\bibfield  {journal} {\bibinfo  {journal}
  {Science}\ }\textbf {\bibinfo {volume} {345}},\ \bibinfo {pages} {424}
  (\bibinfo {year} {2014})}\BibitemShut {NoStop}%
\bibitem [{\citenamefont {T{\'{o}}th}\ and\ \citenamefont
  {Apellaniz}(2014)}]{toth14quantum}%
  \BibitemOpen
  \bibfield  {author} {\bibinfo {author} {\bibfnamefont {G.}~\bibnamefont
  {T{\'{o}}th}}\ and\ \bibinfo {author} {\bibfnamefont {I.}~\bibnamefont
  {Apellaniz}},\ }\href {\doibase 10.1088/1751-8113/47/42/424006} {\bibfield
  {journal} {\bibinfo  {journal} {J. Phys. A: Math. Theor.}\ }\textbf {\bibinfo
  {volume} {47}},\ \bibinfo {pages} {424006} (\bibinfo {year}
  {2014})}\BibitemShut {NoStop}%
\bibitem [{\citenamefont {Dowling}(2008)}]{dowling08quantum}%
  \BibitemOpen
  \bibfield  {author} {\bibinfo {author} {\bibfnamefont {J.~P.}\ \bibnamefont
  {Dowling}},\ }\href {\doibase 10.1080/00107510802091298} {\bibfield
  {journal} {\bibinfo  {journal} {Contemp. Phys.}\ }\textbf {\bibinfo {volume}
  {49}},\ \bibinfo {pages} {125} (\bibinfo {year} {2008})}\BibitemShut
  {NoStop}%
\bibitem [{\citenamefont {Joo}\ \emph {et~al.}(2011)\citenamefont {Joo},
  \citenamefont {Munro},\ and\ \citenamefont {Spiller}}]{joo11quantum}%
  \BibitemOpen
  \bibfield  {author} {\bibinfo {author} {\bibfnamefont {J.}~\bibnamefont
  {Joo}}, \bibinfo {author} {\bibfnamefont {W.~J.}\ \bibnamefont {Munro}}, \
  and\ \bibinfo {author} {\bibfnamefont {T.~P.}\ \bibnamefont {Spiller}},\
  }\href {\doibase 10.1103/PhysRevLett.107.083601} {\bibfield  {journal}
  {\bibinfo  {journal} {Phys. Rev. Lett.}\ }\textbf {\bibinfo {volume} {107}},\
  \bibinfo {pages} {083601} (\bibinfo {year} {2011})}\BibitemShut {NoStop}%
\bibitem [{\citenamefont {Giesbers}\ \emph {et~al.}(2008)\citenamefont
  {Giesbers}, \citenamefont {Rietveld}, \citenamefont {Houtzager},
  \citenamefont {Zeitler}, \citenamefont {Yang}, \citenamefont {Novoselov},
  \citenamefont {Geim},\ and\ \citenamefont {Maan}}]{giesbers08quantum}%
  \BibitemOpen
  \bibfield  {author} {\bibinfo {author} {\bibfnamefont {A.~J.~M.}\
  \bibnamefont {Giesbers}}, \bibinfo {author} {\bibfnamefont {G.}~\bibnamefont
  {Rietveld}}, \bibinfo {author} {\bibfnamefont {E.}~\bibnamefont {Houtzager}},
  \bibinfo {author} {\bibfnamefont {U.}~\bibnamefont {Zeitler}}, \bibinfo
  {author} {\bibfnamefont {R.}~\bibnamefont {Yang}}, \bibinfo {author}
  {\bibfnamefont {K.~S.}\ \bibnamefont {Novoselov}}, \bibinfo {author}
  {\bibfnamefont {A.~K.}\ \bibnamefont {Geim}}, \ and\ \bibinfo {author}
  {\bibfnamefont {J.~C.}\ \bibnamefont {Maan}},\ }\href {\doibase
  10.1063/1.3043426} {\bibfield  {journal} {\bibinfo  {journal} {Appl. Phys.
  Lett.}\ }\textbf {\bibinfo {volume} {93}},\ \bibinfo {pages} {222109}
  (\bibinfo {year} {2008})}\BibitemShut {NoStop}%
\bibitem [{\citenamefont {Zanardi}\ \emph {et~al.}(2008)\citenamefont
  {Zanardi}, \citenamefont {Paris},\ and\ \citenamefont
  {Campos~Venuti}}]{zanardi08quantum}%
  \BibitemOpen
  \bibfield  {author} {\bibinfo {author} {\bibfnamefont {P.}~\bibnamefont
  {Zanardi}}, \bibinfo {author} {\bibfnamefont {M.~G.~A.}\ \bibnamefont
  {Paris}}, \ and\ \bibinfo {author} {\bibfnamefont {L.}~\bibnamefont
  {Campos~Venuti}},\ }\href {\doibase 10.1103/PhysRevA.78.042105} {\bibfield
  {journal} {\bibinfo  {journal} {Phys. Rev. A}\ }\textbf {\bibinfo {volume}
  {78}},\ \bibinfo {pages} {042105} (\bibinfo {year} {2008})}\BibitemShut
  {NoStop}%
\bibitem [{\citenamefont {Gross}(2012)}]{gross12spin}%
  \BibitemOpen
  \bibfield  {author} {\bibinfo {author} {\bibfnamefont {C.}~\bibnamefont
  {Gross}},\ }\href {\doibase 10.1088/0953-4075/45/10/103001} {\bibfield
  {journal} {\bibinfo  {journal} {J. Phys. B: At. Mol. Opt. Phys.}\ }\textbf
  {\bibinfo {volume} {45}},\ \bibinfo {pages} {103001} (\bibinfo {year}
  {2012})}\BibitemShut {NoStop}%
\bibitem [{\citenamefont {Hovhannisyan}\ and\ \citenamefont
  {Correa}(2018)}]{hovhannisyan18measuring}%
  \BibitemOpen
  \bibfield  {author} {\bibinfo {author} {\bibfnamefont {K.~V.}\ \bibnamefont
  {Hovhannisyan}}\ and\ \bibinfo {author} {\bibfnamefont {L.~A.}\ \bibnamefont
  {Correa}},\ }\href {\doibase 10.1103/PhysRevB.98.045101} {\bibfield
  {journal} {\bibinfo  {journal} {Phys. Rev. B}\ }\textbf {\bibinfo {volume}
  {98}},\ \bibinfo {pages} {045101} (\bibinfo {year} {2018})}\BibitemShut
  {NoStop}%
\bibitem [{\citenamefont {Brunelli}\ \emph {et~al.}(2011)\citenamefont
  {Brunelli}, \citenamefont {Olivares},\ and\ \citenamefont
  {Paris}}]{brunelli11}%
  \BibitemOpen
  \bibfield  {author} {\bibinfo {author} {\bibfnamefont {M.}~\bibnamefont
  {Brunelli}}, \bibinfo {author} {\bibfnamefont {S.}~\bibnamefont {Olivares}},
  \ and\ \bibinfo {author} {\bibfnamefont {M.~G.~A.}\ \bibnamefont {Paris}},\
  }\href {\doibase 10.1103/PhysRevA.84.032105} {\bibfield  {journal} {\bibinfo
  {journal} {Phys. Rev. A}\ }\textbf {\bibinfo {volume} {84}},\ \bibinfo
  {pages} {032105} (\bibinfo {year} {2011})}\BibitemShut {NoStop}%
\bibitem [{\citenamefont {Pati}\ \emph {et~al.}(2019)\citenamefont {Pati},
  \citenamefont {Mukhopadhyay}, \citenamefont {Chakraborty},\ and\
  \citenamefont {Ghosh}}]{pati19quantum}%
  \BibitemOpen
  \bibfield  {author} {\bibinfo {author} {\bibfnamefont {A.~K.}\ \bibnamefont
  {Pati}}, \bibinfo {author} {\bibfnamefont {C.}~\bibnamefont {Mukhopadhyay}},
  \bibinfo {author} {\bibfnamefont {S.}~\bibnamefont {Chakraborty}}, \ and\
  \bibinfo {author} {\bibfnamefont {S.}~\bibnamefont {Ghosh}},\ }\href
  {https://arxiv.org/abs/1901.07415} {\bibfield  {journal} {\bibinfo  {journal}
  {arXiv preprint arXiv:1901.07415}\ } (\bibinfo {year} {2019})}\BibitemShut
  {NoStop}%
\bibitem [{\citenamefont {Muessel}\ \emph {et~al.}(2014)\citenamefont
  {Muessel}, \citenamefont {Strobel}, \citenamefont {Linnemann}, \citenamefont
  {Hume},\ and\ \citenamefont {Oberthaler}}]{muessel14scalable}%
  \BibitemOpen
  \bibfield  {author} {\bibinfo {author} {\bibfnamefont {W.}~\bibnamefont
  {Muessel}}, \bibinfo {author} {\bibfnamefont {H.}~\bibnamefont {Strobel}},
  \bibinfo {author} {\bibfnamefont {D.}~\bibnamefont {Linnemann}}, \bibinfo
  {author} {\bibfnamefont {D.~B.}\ \bibnamefont {Hume}}, \ and\ \bibinfo
  {author} {\bibfnamefont {M.~K.}\ \bibnamefont {Oberthaler}},\ }\href
  {\doibase 10.1103/PhysRevLett.113.103004} {\bibfield  {journal} {\bibinfo
  {journal} {Phys. Rev. Lett.}\ }\textbf {\bibinfo {volume} {113}},\ \bibinfo
  {pages} {103004} (\bibinfo {year} {2014})}\BibitemShut {NoStop}%
\bibitem [{\citenamefont {Brask}\ \emph {et~al.}(2015)\citenamefont {Brask},
  \citenamefont {Chaves},\ and\ \citenamefont {Ko\l{}ody\ifmmode~\acute{n}\else
  \'{n}\fi{}ski}}]{brask15improved}%
  \BibitemOpen
  \bibfield  {author} {\bibinfo {author} {\bibfnamefont {J.~B.}\ \bibnamefont
  {Brask}}, \bibinfo {author} {\bibfnamefont {R.}~\bibnamefont {Chaves}}, \
  and\ \bibinfo {author} {\bibfnamefont {J.}~\bibnamefont
  {Ko\l{}ody\ifmmode~\acute{n}\else \'{n}\fi{}ski}},\ }\href {\doibase
  10.1103/PhysRevX.5.031010} {\bibfield  {journal} {\bibinfo  {journal} {Phys.
  Rev. X}\ }\textbf {\bibinfo {volume} {5}},\ \bibinfo {pages} {031010}
  (\bibinfo {year} {2015})}\BibitemShut {NoStop}%
\bibitem [{\citenamefont {Albarelli}\ \emph {et~al.}(2017)\citenamefont
  {Albarelli}, \citenamefont {Rossi}, \citenamefont {Paris},\ and\
  \citenamefont {Genoni}}]{albarelli17ultimate}%
  \BibitemOpen
  \bibfield  {author} {\bibinfo {author} {\bibfnamefont {F.}~\bibnamefont
  {Albarelli}}, \bibinfo {author} {\bibfnamefont {M.~A.~C.}\ \bibnamefont
  {Rossi}}, \bibinfo {author} {\bibfnamefont {M.~G.~A.}\ \bibnamefont {Paris}},
  \ and\ \bibinfo {author} {\bibfnamefont {M.~G.}\ \bibnamefont {Genoni}},\
  }\href {\doibase 10.1088/1367-2630/aa9840} {\bibfield  {journal} {\bibinfo
  {journal} {New J. Phys.}\ }\textbf {\bibinfo {volume} {19}},\ \bibinfo
  {pages} {123011} (\bibinfo {year} {2017})}\BibitemShut {NoStop}%
\bibitem [{\citenamefont {Pezz\'e}\ and\ \citenamefont
  {Smerzi}(2009)}]{pezze09entanglement}%
  \BibitemOpen
  \bibfield  {author} {\bibinfo {author} {\bibfnamefont {L.}~\bibnamefont
  {Pezz\'e}}\ and\ \bibinfo {author} {\bibfnamefont {A.}~\bibnamefont
  {Smerzi}},\ }\href {\doibase 10.1103/PhysRevLett.102.100401} {\bibfield
  {journal} {\bibinfo  {journal} {Phys. Rev. Lett.}\ }\textbf {\bibinfo
  {volume} {102}},\ \bibinfo {pages} {100401} (\bibinfo {year}
  {2009})}\BibitemShut {NoStop}%
\bibitem [{\citenamefont {Demkowicz-Dobrza\'{n}ski}\ \emph
  {et~al.}(2012)\citenamefont {Demkowicz-Dobrza\'{n}ski}, \citenamefont
  {Kolody\'{n}ski},\ and\ \citenamefont {Guta}}]{dobrzanski12the}%
  \BibitemOpen
  \bibfield  {author} {\bibinfo {author} {\bibfnamefont {R.}~\bibnamefont
  {Demkowicz-Dobrza\'{n}ski}}, \bibinfo {author} {\bibfnamefont
  {J.}~\bibnamefont {Kolody\'{n}ski}}, \ and\ \bibinfo {author} {\bibfnamefont
  {M.}~\bibnamefont {Guta}},\ }\href {\doibase 10.1038/ncomms2067} {\bibfield
  {journal} {\bibinfo  {journal} {Nat. Commun.}\ }\textbf {\bibinfo {volume}
  {3}},\ \bibinfo {pages} {1063} (\bibinfo {year} {2012})}\BibitemShut
  {NoStop}%
\bibitem [{\citenamefont {Braunstein}\ and\ \citenamefont
  {Caves}(1994)}]{caves94}%
  \BibitemOpen
  \bibfield  {author} {\bibinfo {author} {\bibfnamefont {S.~L.}\ \bibnamefont
  {Braunstein}}\ and\ \bibinfo {author} {\bibfnamefont {C.~M.}\ \bibnamefont
  {Caves}},\ }\href {\doibase 10.1103/PhysRevLett.72.3439} {\bibfield
  {journal} {\bibinfo  {journal} {Phys. Rev. Lett.}\ }\textbf {\bibinfo
  {volume} {72}},\ \bibinfo {pages} {3439} (\bibinfo {year}
  {1994})}\BibitemShut {NoStop}%
\bibitem [{\citenamefont {Zhang}\ \emph {et~al.}(2013)\citenamefont {Zhang},
  \citenamefont {Li}, \citenamefont {Yang},\ and\ \citenamefont
  {Jin}}]{zhang13}%
  \BibitemOpen
  \bibfield  {author} {\bibinfo {author} {\bibfnamefont {Y.~M.}\ \bibnamefont
  {Zhang}}, \bibinfo {author} {\bibfnamefont {X.~W.}\ \bibnamefont {Li}},
  \bibinfo {author} {\bibfnamefont {W.}~\bibnamefont {Yang}}, \ and\ \bibinfo
  {author} {\bibfnamefont {G.~R.}\ \bibnamefont {Jin}},\ }\href {\doibase
  10.1103/PhysRevA.88.043832} {\bibfield  {journal} {\bibinfo  {journal} {Phys.
  Rev. A}\ }\textbf {\bibinfo {volume} {88}},\ \bibinfo {pages} {043832}
  (\bibinfo {year} {2013})}\BibitemShut {NoStop}%
\bibitem [{\citenamefont {De~Pasquale}\ \emph {et~al.}(2016)\citenamefont
  {De~Pasquale}, \citenamefont {Rossini}, \citenamefont {Fazio},\ and\
  \citenamefont {Giovannetti}}]{pasquale16local}%
  \BibitemOpen
  \bibfield  {author} {\bibinfo {author} {\bibfnamefont {A.}~\bibnamefont
  {De~Pasquale}}, \bibinfo {author} {\bibfnamefont {D.}~\bibnamefont
  {Rossini}}, \bibinfo {author} {\bibfnamefont {R.}~\bibnamefont {Fazio}}, \
  and\ \bibinfo {author} {\bibfnamefont {V.}~\bibnamefont {Giovannetti}},\
  }\href {\doibase 10.1038/ncomms12782} {\bibfield  {journal} {\bibinfo
  {journal} {Nat. Commun.}\ }\textbf {\bibinfo {volume} {7}},\ \bibinfo {pages}
  {12782} (\bibinfo {year} {2016})}\BibitemShut {NoStop}%
\bibitem [{\citenamefont {Gefen}\ \emph {et~al.}(2017)\citenamefont {Gefen},
  \citenamefont {Jelezko},\ and\ \citenamefont {Retzker}}]{gefen17}%
  \BibitemOpen
  \bibfield  {author} {\bibinfo {author} {\bibfnamefont {T.}~\bibnamefont
  {Gefen}}, \bibinfo {author} {\bibfnamefont {F.}~\bibnamefont {Jelezko}}, \
  and\ \bibinfo {author} {\bibfnamefont {A.}~\bibnamefont {Retzker}},\ }\href
  {\doibase 10.1103/PhysRevA.96.032310} {\bibfield  {journal} {\bibinfo
  {journal} {Phys. Rev. A}\ }\textbf {\bibinfo {volume} {96}},\ \bibinfo
  {pages} {032310} (\bibinfo {year} {2017})}\BibitemShut {NoStop}%
\bibitem [{\citenamefont {Mukherjee}\ \emph {et~al.}(2017)\citenamefont
  {Mukherjee}, \citenamefont {Zwick}, \citenamefont {Ghosh}, \citenamefont
  {Chen},\ and\ \citenamefont {Kurizki}}]{mukherjee17enhanced}%
  \BibitemOpen
  \bibfield  {author} {\bibinfo {author} {\bibfnamefont {V.}~\bibnamefont
  {Mukherjee}}, \bibinfo {author} {\bibfnamefont {A.}~\bibnamefont {Zwick}},
  \bibinfo {author} {\bibfnamefont {A.}~\bibnamefont {Ghosh}}, \bibinfo
  {author} {\bibfnamefont {X.}~\bibnamefont {Chen}}, \ and\ \bibinfo {author}
  {\bibfnamefont {G.}~\bibnamefont {Kurizki}},\ }\href
  {\doibase 10.1038/s42005-019-0265-y} {\bibfield  {journal} {\bibinfo  {journal}
  {Commun. Phys.}\ }\textbf {\bibinfo {volume} {2}},\ \bibinfo
{pages} {1--8} (\bibinfo {year} {2019})}\BibitemShut
  {NoStop}%
\bibitem [{\citenamefont {Alicki}(1979)}]{alicki1979quantum}%
  \BibitemOpen
  \bibfield  {author} {\bibinfo {author} {\bibfnamefont {R.}~\bibnamefont
  {Alicki}},\ }\href {\doibase 10.1088/0305-4470/12/5/007} {\bibfield
  {journal} {\bibinfo  {journal} {J. Phys. A}\ }\textbf {\bibinfo {volume}
  {12}},\ \bibinfo {pages} {L103} (\bibinfo {year} {1979})}\BibitemShut
  {NoStop}%
\bibitem [{\citenamefont {Scully}\ \emph {et~al.}(2003)\citenamefont {Scully},
  \citenamefont {Zubairy}, \citenamefont {Agarwal},\ and\ \citenamefont
  {Walther}}]{scully03extracting}%
  \BibitemOpen
  \bibfield  {author} {\bibinfo {author} {\bibfnamefont {M.~O.}\ \bibnamefont
  {Scully}}, \bibinfo {author} {\bibfnamefont {M.~S.}\ \bibnamefont {Zubairy}},
  \bibinfo {author} {\bibfnamefont {G.~S.}\ \bibnamefont {Agarwal}}, \ and\
  \bibinfo {author} {\bibfnamefont {H.}~\bibnamefont {Walther}},\ }\href
  {\doibase 10.1126/science.1078955} {\bibfield  {journal} {\bibinfo  {journal}
  {Science}\ }\textbf {\bibinfo {volume} {299}},\ \bibinfo {pages} {862}
  (\bibinfo {year} {2003})}\BibitemShut {NoStop}%
\bibitem [{\citenamefont {Kosloff}(2013)}]{kosloff13quantum}%
  \BibitemOpen
  \bibfield  {author} {\bibinfo {author} {\bibfnamefont {R.}~\bibnamefont
  {Kosloff}},\ }\href {\doibase 10.3390/e15062100} {\bibfield  {journal}
  {\bibinfo  {journal} {Entropy}\ }\textbf {\bibinfo {volume} {15}},\ \bibinfo
  {pages} {2100} (\bibinfo {year} {2013})}\BibitemShut {NoStop}%
\bibitem [{\citenamefont {Zhang}\ \emph {et~al.}(2014)\citenamefont {Zhang},
  \citenamefont {Bariani},\ and\ \citenamefont {Meystre}}]{zhang14quantum}%
  \BibitemOpen
  \bibfield  {author} {\bibinfo {author} {\bibfnamefont {K.}~\bibnamefont
  {Zhang}}, \bibinfo {author} {\bibfnamefont {F.}~\bibnamefont {Bariani}}, \
  and\ \bibinfo {author} {\bibfnamefont {P.}~\bibnamefont {Meystre}},\ }\href
  {\doibase 10.1103/PhysRevLett.112.150602} {\bibfield  {journal} {\bibinfo
  {journal} {Phys. Rev. Lett.}\ }\textbf {\bibinfo {volume} {112}},\ \bibinfo
  {pages} {150602} (\bibinfo {year} {2014})}\BibitemShut {NoStop}%
\bibitem [{\citenamefont {Gelbwaser-Klimovsky}\ \emph
  {et~al.}(2015)\citenamefont {Gelbwaser-Klimovsky}, \citenamefont {Niedenzu},\
  and\ \citenamefont {Kurizki}}]{klimovsky15thermodynamics}%
  \BibitemOpen
  \bibfield  {author} {\bibinfo {author} {\bibfnamefont {D.}~\bibnamefont
  {Gelbwaser-Klimovsky}}, \bibinfo {author} {\bibfnamefont {W.}~\bibnamefont
  {Niedenzu}}, \ and\ \bibinfo {author} {\bibfnamefont {G.}~\bibnamefont
  {Kurizki}},\ }\href {\doibase 10.1016/bs.aamop.2015.07.002} {\bibfield
  {journal} {\bibinfo  {journal} {Adv. At. Mol. Opt. Phys.}\ }\textbf {\bibinfo
  {volume} {64}},\ \bibinfo {pages} {329} (\bibinfo {year} {2015})}\BibitemShut
  {NoStop}%
\bibitem [{\citenamefont {Campisi}\ and\ \citenamefont
  {Fazio}(2016{\natexlab{a}})}]{campisi16the}%
  \BibitemOpen
  \bibfield  {author} {\bibinfo {author} {\bibfnamefont {M.}~\bibnamefont
  {Campisi}}\ and\ \bibinfo {author} {\bibfnamefont {R.}~\bibnamefont
  {Fazio}},\ }\href {\doibase 10.1038/ncomms11895} {\bibfield  {journal}
  {\bibinfo  {journal} {Nat. Comm.}\ }\textbf {\bibinfo {volume} {7}},\
  \bibinfo {pages} {11895} (\bibinfo {year} {2016}{\natexlab{a}})}\BibitemShut
  {NoStop}%
\bibitem [{\citenamefont {Vinjanampathy}\ and\ \citenamefont
  {Anders}(2016)}]{vinjanampathy16quantum}%
  \BibitemOpen
  \bibfield  {author} {\bibinfo {author} {\bibfnamefont {S.}~\bibnamefont
  {Vinjanampathy}}\ and\ \bibinfo {author} {\bibfnamefont {J.}~\bibnamefont
  {Anders}},\ }\href {\doibase 10.1080/00107514.2016.1201896} {\bibfield
  {journal} {\bibinfo  {journal} {Contemp. Phys.}\ }\textbf {\bibinfo {volume}
  {57}},\ \bibinfo {pages} {1} (\bibinfo {year} {2016})}\BibitemShut {NoStop}%
\bibitem [{\citenamefont {Binder}\ \emph {et~al.}(2019)\citenamefont {Binder},
  \citenamefont {Correa}, \citenamefont {Gogolin}, \citenamefont {Anders},\
  and\ \citenamefont {Adesso}}]{binder2019thermodynamicsbook}%
  \BibitemOpen
  \bibinfo {editor} {\bibfnamefont {F.}~\bibnamefont {Binder}}, \bibinfo
  {editor} {\bibfnamefont {L.~A.}\ \bibnamefont {Correa}}, \bibinfo {editor}
  {\bibfnamefont {C.}~\bibnamefont {Gogolin}}, \bibinfo {editor} {\bibfnamefont
  {J.}~\bibnamefont {Anders}}, \ and\ \bibinfo {editor} {\bibfnamefont
  {G.}~\bibnamefont {Adesso}},\ eds.,\ \href {\doibase
  10.1007/978-3-319-99046-0} {\emph {\bibinfo {title} {Thermodynamics in the
  Quantum Regime}}}\ (\bibinfo  {publisher} {Springer-Verlag Cham},\ \bibinfo
  {year} {2019})\BibitemShut {NoStop}%
\bibitem [{\citenamefont {Hofer}\ \emph {et~al.}(2017)\citenamefont {Hofer},
  \citenamefont {Brask}, \citenamefont {Perarnau-Llobet},\ and\ \citenamefont
  {Brunner}}]{brunner17quantum}%
  \BibitemOpen
  \bibfield  {author} {\bibinfo {author} {\bibfnamefont {P.~P.}\ \bibnamefont
  {Hofer}}, \bibinfo {author} {\bibfnamefont {J.~B.}\ \bibnamefont {Brask}},
  \bibinfo {author} {\bibfnamefont {M.}~\bibnamefont {Perarnau-Llobet}}, \ and\
  \bibinfo {author} {\bibfnamefont {N.}~\bibnamefont {Brunner}},\ }\href
  {\doibase 10.1103/PhysRevLett.119.090603} {\bibfield  {journal} {\bibinfo
  {journal} {Phys. Rev. Lett.}\ }\textbf {\bibinfo {volume} {119}},\ \bibinfo
  {pages} {090603} (\bibinfo {year} {2017})}\BibitemShut {NoStop}%
\bibitem [{\citenamefont {Uzdin}\ \emph {et~al.}(2015)\citenamefont {Uzdin},
  \citenamefont {Levy},\ and\ \citenamefont {Kosloff}}]{uzdin15equivalence}%
  \BibitemOpen
  \bibfield  {author} {\bibinfo {author} {\bibfnamefont {R.}~\bibnamefont
  {Uzdin}}, \bibinfo {author} {\bibfnamefont {A.}~\bibnamefont {Levy}}, \ and\
  \bibinfo {author} {\bibfnamefont {R.}~\bibnamefont {Kosloff}},\ }\href
  {\doibase 10.1103/PhysRevX.5.031044} {\bibfield  {journal} {\bibinfo
  {journal} {Phys. Rev. X}\ }\textbf {\bibinfo {volume} {5}},\ \bibinfo {pages}
  {031044} (\bibinfo {year} {2015})}\BibitemShut {NoStop}%
\bibitem [{\citenamefont {Campisi}\ \emph {et~al.}(2015)\citenamefont
  {Campisi}, \citenamefont {Pekola},\ and\ \citenamefont
  {Fazio}}]{campisi2015nonequilibrium}%
  \BibitemOpen
  \bibfield  {author} {\bibinfo {author} {\bibfnamefont {M.}~\bibnamefont
  {Campisi}}, \bibinfo {author} {\bibfnamefont {J.}~\bibnamefont {Pekola}}, \
  and\ \bibinfo {author} {\bibfnamefont {R.}~\bibnamefont {Fazio}},\ }\href
  {\doibase 10.1088/1367-2630/17/3/035012} {\bibfield  {journal} {\bibinfo
  {journal} {New J. Phys.}\ }\textbf {\bibinfo {volume} {17}},\ \bibinfo
  {pages} {035012} (\bibinfo {year} {2015})}\BibitemShut {NoStop}%
\bibitem [{\citenamefont {Campisi}\ and\ \citenamefont
  {Fazio}(2016{\natexlab{b}})}]{campisi2016dissipation}%
  \BibitemOpen
  \bibfield  {author} {\bibinfo {author} {\bibfnamefont {M.}~\bibnamefont
  {Campisi}}\ and\ \bibinfo {author} {\bibfnamefont {R.}~\bibnamefont
  {Fazio}},\ }\href {\doibase 10.1088/1751-8113/49/34/345002} {\bibfield
  {journal} {\bibinfo  {journal} {J. Phys. A: Math. Theor.}\ }\textbf {\bibinfo
  {volume} {49}},\ \bibinfo {pages} {345002} (\bibinfo {year}
  {2016}{\natexlab{b}})}\BibitemShut {NoStop}%
\bibitem [{\citenamefont {Gelbwaser-Klimovsky}\ \emph
  {et~al.}(2013)\citenamefont {Gelbwaser-Klimovsky}, \citenamefont {Alicki},\
  and\ \citenamefont {Kurizki}}]{klimovsky13minimal}%
  \BibitemOpen
  \bibfield  {author} {\bibinfo {author} {\bibfnamefont {D.}~\bibnamefont
  {Gelbwaser-Klimovsky}}, \bibinfo {author} {\bibfnamefont {R.}~\bibnamefont
  {Alicki}}, \ and\ \bibinfo {author} {\bibfnamefont {G.}~\bibnamefont
  {Kurizki}},\ }\href {\doibase 10.1103/PhysRevE.87.012140} {\bibfield
  {journal} {\bibinfo  {journal} {Phys. Rev. E}\ }\textbf {\bibinfo {volume}
  {87}},\ \bibinfo {pages} {012140} (\bibinfo {year} {2013})}\BibitemShut
  {NoStop}%
\bibitem [{\citenamefont {Ern\'e}\ \emph {et~al.}(1976)\citenamefont {Ern\'e},
  \citenamefont {Hahlbohm},\ and\ \citenamefont {L\"ubbig}}]{erne76theory}%
  \BibitemOpen
  \bibfield  {author} {\bibinfo {author} {\bibfnamefont {S.~N.}\ \bibnamefont
  {Ern\'e}}, \bibinfo {author} {\bibfnamefont {H.-D.}\ \bibnamefont
  {Hahlbohm}}, \ and\ \bibinfo {author} {\bibfnamefont {H.}~\bibnamefont
  {L\"ubbig}},\ }\href {\doibase 10.1063/1.322574} {\bibfield  {journal}
  {\bibinfo  {journal} {J. Appl. Phys.}\ }\textbf {\bibinfo {volume} {47}},\
  \bibinfo {pages} {5440} (\bibinfo {year} {1976})}\BibitemShut {NoStop}%
\bibitem [{\citenamefont {Cleuziou}\ \emph {et~al.}(2006)\citenamefont
  {Cleuziou}, \citenamefont {Wernsdorfer}, \citenamefont {Bouchiat},
  \citenamefont {Ondar{\c{c}}uhu},\ and\ \citenamefont
  {Monthioux}}]{cleuziou06carbon}%
  \BibitemOpen
  \bibfield  {author} {\bibinfo {author} {\bibfnamefont {J.-P.}\ \bibnamefont
  {Cleuziou}}, \bibinfo {author} {\bibfnamefont {W.}~\bibnamefont
  {Wernsdorfer}}, \bibinfo {author} {\bibfnamefont {V.}~\bibnamefont
  {Bouchiat}}, \bibinfo {author} {\bibfnamefont {T.}~\bibnamefont
  {Ondar{\c{c}}uhu}}, \ and\ \bibinfo {author} {\bibfnamefont {M.}~\bibnamefont
  {Monthioux}},\ }\href {\doibase 10.1038/nnano.2006.54} {\bibfield  {journal}
  {\bibinfo  {journal} {Nat. Nanotechnol.}\ }\textbf {\bibinfo {volume} {1}},\
  \bibinfo {pages} {53} (\bibinfo {year} {2006})}\BibitemShut {NoStop}%
\bibitem [{\citenamefont {Buffoni}\ \emph {et~al.}(2019)\citenamefont
  {Buffoni}, \citenamefont {Solfanelli}, \citenamefont {Verrucchi},
  \citenamefont {Cuccoli},\ and\ \citenamefont {Campisi}}]{buffoni2019quantum}%
  \BibitemOpen
  \bibfield  {author} {\bibinfo {author} {\bibfnamefont {L.}~\bibnamefont
  {Buffoni}}, \bibinfo {author} {\bibfnamefont {A.}~\bibnamefont {Solfanelli}},
  \bibinfo {author} {\bibfnamefont {P.}~\bibnamefont {Verrucchi}}, \bibinfo
  {author} {\bibfnamefont {A.}~\bibnamefont {Cuccoli}}, \ and\ \bibinfo
  {author} {\bibfnamefont {M.}~\bibnamefont {Campisi}},\ }\href {\doibase
  10.1103/PhysRevLett.122.070603} {\bibfield  {journal} {\bibinfo  {journal}
  {Phys. Rev. Lett.}\ }\textbf {\bibinfo {volume} {122}},\ \bibinfo {pages}
  {070603} (\bibinfo {year} {2019})}\BibitemShut {NoStop}%
\bibitem [{\citenamefont {Mukherjee}\ \emph {et~al.}(2016)\citenamefont
  {Mukherjee}, \citenamefont {Niedenzu}, \citenamefont {Kofman},\ and\
  \citenamefont {Kurizki}}]{mukherjee16speed}%
  \BibitemOpen
  \bibfield  {author} {\bibinfo {author} {\bibfnamefont {V.}~\bibnamefont
  {Mukherjee}}, \bibinfo {author} {\bibfnamefont {W.}~\bibnamefont {Niedenzu}},
  \bibinfo {author} {\bibfnamefont {A.~G.}\ \bibnamefont {Kofman}}, \ and\
  \bibinfo {author} {\bibfnamefont {G.}~\bibnamefont {Kurizki}},\ }\href
  {\doibase 10.1103/PhysRevE.94.062109} {\bibfield  {journal} {\bibinfo
  {journal} {Phys. Rev. E}\ }\textbf {\bibinfo {volume} {94}},\ \bibinfo
  {pages} {062109} (\bibinfo {year} {2016})}\BibitemShut {NoStop}%
\bibitem [{\citenamefont {Iftikhar}\ \emph {et~al.}(2016)\citenamefont
  {Iftikhar}, \citenamefont {Anthore}, \citenamefont {Jezouin}, \citenamefont
  {Parmentier}, \citenamefont {Jin}, \citenamefont {Cavanna}, \citenamefont
  {Ouerghi}, \citenamefont {Gennser},\ and\ \citenamefont
  {Pierre}}]{iftikhar16primary}%
  \BibitemOpen
  \bibfield  {author} {\bibinfo {author} {\bibfnamefont {Z.}~\bibnamefont
  {Iftikhar}}, \bibinfo {author} {\bibfnamefont {A.}~\bibnamefont {Anthore}},
  \bibinfo {author} {\bibfnamefont {S.}~\bibnamefont {Jezouin}}, \bibinfo
  {author} {\bibfnamefont {F.~D.}\ \bibnamefont {Parmentier}}, \bibinfo
  {author} {\bibfnamefont {Y.}~\bibnamefont {Jin}}, \bibinfo {author}
  {\bibfnamefont {A.}~\bibnamefont {Cavanna}}, \bibinfo {author} {\bibfnamefont
  {A.}~\bibnamefont {Ouerghi}}, \bibinfo {author} {\bibfnamefont
  {U.}~\bibnamefont {Gennser}}, \ and\ \bibinfo {author} {\bibfnamefont
  {F.}~\bibnamefont {Pierre}},\ }\href {\doibase 10.1038/ncomms12908}
  {\bibfield  {journal} {\bibinfo  {journal} {Nat. Commun.}\ }\textbf {\bibinfo
  {volume} {7}},\ \bibinfo {pages} {12908} (\bibinfo {year}
  {2016})}\BibitemShut {NoStop}%
\bibitem [{\citenamefont {Klaers}\ \emph {et~al.}(2017)\citenamefont {Klaers},
  \citenamefont {Faelt}, \citenamefont {Imamoglu},\ and\ \citenamefont
  {Togan}}]{klaers17squeezed}%
  \BibitemOpen
  \bibfield  {author} {\bibinfo {author} {\bibfnamefont {J.}~\bibnamefont
  {Klaers}}, \bibinfo {author} {\bibfnamefont {S.}~\bibnamefont {Faelt}},
  \bibinfo {author} {\bibfnamefont {A.}~\bibnamefont {Imamoglu}}, \ and\
  \bibinfo {author} {\bibfnamefont {E.}~\bibnamefont {Togan}},\ }\href
  {\doibase 10.1103/PhysRevX.7.031044} {\bibfield  {journal} {\bibinfo
  {journal} {Phys. Rev. X}\ }\textbf {\bibinfo {volume} {7}},\ \bibinfo {pages}
  {031044} (\bibinfo {year} {2017})}\BibitemShut {NoStop}%
\bibitem [{\citenamefont {Wingert}\ \emph {et~al.}(2012)\citenamefont
  {Wingert}, \citenamefont {Chen}, \citenamefont {Kwon}, \citenamefont
  {Xiang},\ and\ \citenamefont {Chen}}]{wingert2012ultra}%
  \BibitemOpen
  \bibfield  {author} {\bibinfo {author} {\bibfnamefont {M.~C.}\ \bibnamefont
  {Wingert}}, \bibinfo {author} {\bibfnamefont {Z.~C.~Y.}\ \bibnamefont
  {Chen}}, \bibinfo {author} {\bibfnamefont {S.}~\bibnamefont {Kwon}}, \bibinfo
  {author} {\bibfnamefont {J.}~\bibnamefont {Xiang}}, \ and\ \bibinfo {author}
  {\bibfnamefont {R.}~\bibnamefont {Chen}},\ }\href {\doibase
  10.1063/1.3681255} {\bibfield  {journal} {\bibinfo  {journal} {Rev. Sci.
  Instrum.}\ }\textbf {\bibinfo {volume} {83}},\ \bibinfo {pages} {024901}
  (\bibinfo {year} {2012})}\BibitemShut {NoStop}%
\bibitem [{\citenamefont {Sangouard}\ \emph {et~al.}(2005)\citenamefont
  {Sangouard}, \citenamefont {Lacour}, \citenamefont {Gu\'erin},\ and\
  \citenamefont {Jauslin}}]{sangouard05fast}%
  \BibitemOpen
  \bibfield  {author} {\bibinfo {author} {\bibfnamefont {N.}~\bibnamefont
  {Sangouard}}, \bibinfo {author} {\bibfnamefont {X.}~\bibnamefont {Lacour}},
  \bibinfo {author} {\bibfnamefont {S.}~\bibnamefont {Gu\'erin}}, \ and\
  \bibinfo {author} {\bibfnamefont {H.~R.}\ \bibnamefont {Jauslin}},\ }\href
  {\doibase 10.1103/PhysRevA.72.062309} {\bibfield  {journal} {\bibinfo
  {journal} {Phys. Rev. A}\ }\textbf {\bibinfo {volume} {72}},\ \bibinfo
  {pages} {062309} (\bibinfo {year} {2005})}\BibitemShut {NoStop}%
\bibitem [{\citenamefont {Liang}\ and\ \citenamefont
  {Li}(2005)}]{liang05realization}%
  \BibitemOpen
  \bibfield  {author} {\bibinfo {author} {\bibfnamefont {L.-m.}\ \bibnamefont
  {Liang}}\ and\ \bibinfo {author} {\bibfnamefont {C.-z.}\ \bibnamefont {Li}},\
  }\href {\doibase 10.1103/PhysRevA.72.024303} {\bibfield  {journal} {\bibinfo
  {journal} {Phys. Rev. A}\ }\textbf {\bibinfo {volume} {72}},\ \bibinfo
  {pages} {024303} (\bibinfo {year} {2005})}\BibitemShut {NoStop}%
\bibitem [{\citenamefont {Zhang}(2007)}]{zhang07mutual}%
  \BibitemOpen
  \bibfield  {author} {\bibinfo {author} {\bibfnamefont {G.-F.}\ \bibnamefont
  {Zhang}},\ }\href {\doibase 10.1088/0953-8984/19/45/456205} {\bibfield
  {journal} {\bibinfo  {journal} {J. Phys.: Condens. Matter}\ }\textbf
  {\bibinfo {volume} {19}},\ \bibinfo {pages} {456205} (\bibinfo {year}
  {2007})}\BibitemShut {NoStop}%
\bibitem [{\citenamefont {Kosloff}\ and\ \citenamefont
  {Rezek}(2017)}]{kosloff17the}%
  \BibitemOpen
  \bibfield  {author} {\bibinfo {author} {\bibfnamefont {R.}~\bibnamefont
  {Kosloff}}\ and\ \bibinfo {author} {\bibfnamefont {Y.}~\bibnamefont
  {Rezek}},\ }\href {\doibase 10.3390/e19040136} {\bibfield  {journal}
  {\bibinfo  {journal} {Entropy}\ }\textbf {\bibinfo {volume} {19}},\ \bibinfo
  {pages} {136} (\bibinfo {year} {2017})}\BibitemShut {NoStop}%
\bibitem [{\citenamefont {Kosloff}\ and\ \citenamefont
  {Levy}(2014)}]{kosloff14quantum}%
  \BibitemOpen
  \bibfield  {author} {\bibinfo {author} {\bibfnamefont {R.}~\bibnamefont
  {Kosloff}}\ and\ \bibinfo {author} {\bibfnamefont {A.}~\bibnamefont {Levy}},\
  }\href {\doibase 10.1146/annurev-physchem-040513-103724} {\bibfield
  {journal} {\bibinfo  {journal} {Annu. Rev. Phys. Chem.}\ }\textbf {\bibinfo
  {volume} {65}},\ \bibinfo {pages} {365} (\bibinfo {year} {2014})}\BibitemShut
  {NoStop}%
\bibitem [{\citenamefont {Breuer}\ and\ \citenamefont
  {Petruccione}(2002)}]{breuerbook}%
  \BibitemOpen
  \bibfield  {author} {\bibinfo {author} {\bibfnamefont {H.-P.}\ \bibnamefont
  {Breuer}}\ and\ \bibinfo {author} {\bibfnamefont {F.}~\bibnamefont
  {Petruccione}},\ }\href@noop {} {\emph {\bibinfo {title} {The Theory of Open
  Quantum Systems}}}\ (\bibinfo  {publisher} {Oxford University Press},\
  \bibinfo {address} {Oxford},\ \bibinfo {year} {2002})\BibitemShut {NoStop}%
\bibitem [{\citenamefont {Safavi-Naeini}\ \emph {et~al.}(2012)\citenamefont
  {Safavi-Naeini}, \citenamefont {Chan}, \citenamefont {Hill}, \citenamefont
  {Alegre}, \citenamefont {Krause},\ and\ \citenamefont {Painter}}]{safavi12}%
  \BibitemOpen
  \bibfield  {author} {\bibinfo {author} {\bibfnamefont {A.~H.}\ \bibnamefont
  {Safavi-Naeini}}, \bibinfo {author} {\bibfnamefont {J.}~\bibnamefont {Chan}},
  \bibinfo {author} {\bibfnamefont {J.~T.}\ \bibnamefont {Hill}}, \bibinfo
  {author} {\bibfnamefont {T.~P.~M.}\ \bibnamefont {Alegre}}, \bibinfo {author}
  {\bibfnamefont {A.}~\bibnamefont {Krause}}, \ and\ \bibinfo {author}
  {\bibfnamefont {O.}~\bibnamefont {Painter}},\ }\href {\doibase
  10.1103/PhysRevLett.108.033602} {\bibfield  {journal} {\bibinfo  {journal}
  {Phys. Rev. Lett.}\ }\textbf {\bibinfo {volume} {108}},\ \bibinfo {pages}
  {033602} (\bibinfo {year} {2012})}\BibitemShut {NoStop}%
\bibitem [{\citenamefont {Hopper}\ \emph {et~al.}(2018)\citenamefont {Hopper},
  \citenamefont {Shulevitz},\ and\ \citenamefont {Bassett}}]{hopper18}%
  \BibitemOpen
  \bibfield  {author} {\bibinfo {author} {\bibfnamefont {D.~A.}\ \bibnamefont
  {Hopper}}, \bibinfo {author} {\bibfnamefont {H.~J.}\ \bibnamefont
  {Shulevitz}}, \ and\ \bibinfo {author} {\bibfnamefont {L.~C.}\ \bibnamefont
  {Bassett}},\ }\href {\doibase 10.3390/mi9090437} {\bibfield  {journal}
  {\bibinfo  {journal} {Micromachines}\ }\textbf {\bibinfo {volume} {9}},\
  \bibinfo {pages} {437} (\bibinfo {year} {2018})}\BibitemShut {NoStop}%
\bibitem [{\citenamefont {Tran}\ \emph {et~al.}(2019)\citenamefont {Tran},
  \citenamefont {Regan}, \citenamefont {Ekimov}, \citenamefont {Mu},
  \citenamefont {Zhou}, \citenamefont {Gao}, \citenamefont {Narang},
  \citenamefont {Solntsev}, \citenamefont {Toth}, \citenamefont {Aharonovich},\
  and\ \citenamefont {Bradac}}]{tran2019anti}%
  \BibitemOpen
  \bibfield  {author} {\bibinfo {author} {\bibfnamefont {T.~T.}\ \bibnamefont
  {Tran}}, \bibinfo {author} {\bibfnamefont {B.}~\bibnamefont {Regan}},
  \bibinfo {author} {\bibfnamefont {E.~A.}\ \bibnamefont {Ekimov}}, \bibinfo
  {author} {\bibfnamefont {Z.}~\bibnamefont {Mu}}, \bibinfo {author}
  {\bibfnamefont {Y.}~\bibnamefont {Zhou}}, \bibinfo {author} {\bibfnamefont
  {W.-b.}\ \bibnamefont {Gao}}, \bibinfo {author} {\bibfnamefont
  {P.}~\bibnamefont {Narang}}, \bibinfo {author} {\bibfnamefont {A.~S.}\
  \bibnamefont {Solntsev}}, \bibinfo {author} {\bibfnamefont {M.}~\bibnamefont
  {Toth}}, \bibinfo {author} {\bibfnamefont {I.}~\bibnamefont {Aharonovich}}, \
  and\ \bibinfo {author} {\bibfnamefont {C.}~\bibnamefont {Bradac}},\ }\href
  {\doibase 10.1126/sciadv.aav9180} {\bibfield  {journal} {\bibinfo  {journal}
  {Sci. Adv.}\ }\textbf {\bibinfo {volume} {5}},\ \bibinfo {pages} {eaav9180}
  (\bibinfo {year} {2019})}\BibitemShut {NoStop}%
\bibitem [{\citenamefont {Vion}\ \emph {et~al.}(2002)\citenamefont {Vion},
  \citenamefont {Aassime}, \citenamefont {Cottet}, \citenamefont {Joyez},
  \citenamefont {Pothier}, \citenamefont {Urbina}, \citenamefont {Esteve},\
  and\ \citenamefont {Devoret}}]{vion02manipulating}%
  \BibitemOpen
  \bibfield  {author} {\bibinfo {author} {\bibfnamefont {D.}~\bibnamefont
  {Vion}}, \bibinfo {author} {\bibfnamefont {A.}~\bibnamefont {Aassime}},
  \bibinfo {author} {\bibfnamefont {A.}~\bibnamefont {Cottet}}, \bibinfo
  {author} {\bibfnamefont {P.}~\bibnamefont {Joyez}}, \bibinfo {author}
  {\bibfnamefont {H.}~\bibnamefont {Pothier}}, \bibinfo {author} {\bibfnamefont
  {C.}~\bibnamefont {Urbina}}, \bibinfo {author} {\bibfnamefont
  {D.}~\bibnamefont {Esteve}}, \ and\ \bibinfo {author} {\bibfnamefont {M.~H.}\
  \bibnamefont {Devoret}},\ }\href {\doibase 10.1126/science.1069372}
  {\bibfield  {journal} {\bibinfo  {journal} {Science}\ }\textbf {\bibinfo
  {volume} {296}},\ \bibinfo {pages} {886} (\bibinfo {year}
  {2002})}\BibitemShut {NoStop}%
\bibitem [{\citenamefont {Wallraff}\ \emph {et~al.}(2004)\citenamefont
  {Wallraff}, \citenamefont {Schuster}, \citenamefont {Blais}, \citenamefont
  {Frunzio}, \citenamefont {Huang}, \citenamefont {Majer}, \citenamefont
  {Kumar}, \citenamefont {Girvin},\ and\ \citenamefont
  {Schoelkopf}}]{wallraff04strong}%
  \BibitemOpen
  \bibfield  {author} {\bibinfo {author} {\bibfnamefont {A.}~\bibnamefont
  {Wallraff}}, \bibinfo {author} {\bibfnamefont {D.~I.}\ \bibnamefont
  {Schuster}}, \bibinfo {author} {\bibfnamefont {A.}~\bibnamefont {Blais}},
  \bibinfo {author} {\bibfnamefont {L.}~\bibnamefont {Frunzio}}, \bibinfo
  {author} {\bibfnamefont {R.~S.}\ \bibnamefont {Huang}}, \bibinfo {author}
  {\bibfnamefont {J.}~\bibnamefont {Majer}}, \bibinfo {author} {\bibfnamefont
  {S.}~\bibnamefont {Kumar}}, \bibinfo {author} {\bibfnamefont {S.~M.}\
  \bibnamefont {Girvin}}, \ and\ \bibinfo {author} {\bibfnamefont {R.~J.}\
  \bibnamefont {Schoelkopf}},\ }\href {\doibase 10.1038/nature02851} {\bibfield
   {journal} {\bibinfo  {journal} {Nature}\ }\textbf {\bibinfo {volume}
  {431}},\ \bibinfo {pages} {162} (\bibinfo {year} {2004})}\BibitemShut
  {NoStop}%
\bibitem [{\citenamefont {Koch}\ \emph {et~al.}(2007)\citenamefont {Koch},
  \citenamefont {Yu}, \citenamefont {Gambetta}, \citenamefont {Houck},
  \citenamefont {Schuster}, \citenamefont {Majer}, \citenamefont {Blais},
  \citenamefont {Devoret}, \citenamefont {Girvin},\ and\ \citenamefont
  {Schoelkopf}}]{koch07charge}%
  \BibitemOpen
  \bibfield  {author} {\bibinfo {author} {\bibfnamefont {J.}~\bibnamefont
  {Koch}}, \bibinfo {author} {\bibfnamefont {T.~M.}\ \bibnamefont {Yu}},
  \bibinfo {author} {\bibfnamefont {J.}~\bibnamefont {Gambetta}}, \bibinfo
  {author} {\bibfnamefont {A.~A.}\ \bibnamefont {Houck}}, \bibinfo {author}
  {\bibfnamefont {D.~I.}\ \bibnamefont {Schuster}}, \bibinfo {author}
  {\bibfnamefont {J.}~\bibnamefont {Majer}}, \bibinfo {author} {\bibfnamefont
  {A.}~\bibnamefont {Blais}}, \bibinfo {author} {\bibfnamefont {M.~H.}\
  \bibnamefont {Devoret}}, \bibinfo {author} {\bibfnamefont {S.~M.}\
  \bibnamefont {Girvin}}, \ and\ \bibinfo {author} {\bibfnamefont {R.~J.}\
  \bibnamefont {Schoelkopf}},\ }\href {\doibase 10.1103/PhysRevA.76.042319}
  {\bibfield  {journal} {\bibinfo  {journal} {Phys. Rev. A}\ }\textbf {\bibinfo
  {volume} {76}},\ \bibinfo {pages} {042319} (\bibinfo {year}
  {2007})}\BibitemShut {NoStop}%
\bibitem [{\citenamefont {Grimm}\ \emph {et~al.}(2000)\citenamefont {Grimm},
  \citenamefont {Weidem\"uller},\ and\ \citenamefont
  {Ovchinnikov}}]{grimm2000optical}%
  \BibitemOpen
  \bibfield  {author} {\bibinfo {author} {\bibfnamefont {R.}~\bibnamefont
  {Grimm}}, \bibinfo {author} {\bibfnamefont {M.}~\bibnamefont
  {Weidem\"uller}}, \ and\ \bibinfo {author} {\bibfnamefont {Y.~B.}\
  \bibnamefont {Ovchinnikov}},\ }\href {\doibase 10.1016/S1049-250X(08)60186-X}
  {\bibfield  {journal} {\bibinfo  {journal} {Adv. At. Mol. Opt. Phys.}\
  }\textbf {\bibinfo {volume} {42}},\ \bibinfo {pages} {95 } (\bibinfo {year}
  {2000})}\BibitemShut {NoStop}%
\bibitem [{\citenamefont {Schindler}\ \emph {et~al.}(2013)\citenamefont
  {Schindler}, \citenamefont {Nigg}, \citenamefont {Monz}, \citenamefont
  {Barreiro}, \citenamefont {Martinez}, \citenamefont {Wang}, \citenamefont
  {Quint}, \citenamefont {Brandl}, \citenamefont {Nebendahl}, \citenamefont
  {Roos}, \citenamefont {Chwalla}, \citenamefont {Hennrich},\ and\
  \citenamefont {Blatt}}]{schindler2013quantum}%
  \BibitemOpen
  \bibfield  {author} {\bibinfo {author} {\bibfnamefont {P.}~\bibnamefont
  {Schindler}}, \bibinfo {author} {\bibfnamefont {D.}~\bibnamefont {Nigg}},
  \bibinfo {author} {\bibfnamefont {T.}~\bibnamefont {Monz}}, \bibinfo {author}
  {\bibfnamefont {J.~T.}\ \bibnamefont {Barreiro}}, \bibinfo {author}
  {\bibfnamefont {E.}~\bibnamefont {Martinez}}, \bibinfo {author}
  {\bibfnamefont {S.~X.}\ \bibnamefont {Wang}}, \bibinfo {author}
  {\bibfnamefont {S.}~\bibnamefont {Quint}}, \bibinfo {author} {\bibfnamefont
  {M.~F.}\ \bibnamefont {Brandl}}, \bibinfo {author} {\bibfnamefont
  {V.}~\bibnamefont {Nebendahl}}, \bibinfo {author} {\bibfnamefont {C.~F.}\
  \bibnamefont {Roos}}, \bibinfo {author} {\bibfnamefont {M.}~\bibnamefont
  {Chwalla}}, \bibinfo {author} {\bibfnamefont {M.}~\bibnamefont {Hennrich}}, \
  and\ \bibinfo {author} {\bibfnamefont {R.}~\bibnamefont {Blatt}},\ }\href
  {\doibase 10.1088/1367-2630/15/12/123012} {\bibfield  {journal} {\bibinfo
  {journal} {New J. Phys.}\ }\textbf {\bibinfo {volume} {15}},\ \bibinfo
  {pages} {123012} (\bibinfo {year} {2013})}\BibitemShut {NoStop}%
\bibitem [{\citenamefont {von Lindenfels}\ \emph {et~al.}(2019)\citenamefont
  {von Lindenfels}, \citenamefont {Gr\"ab}, \citenamefont {Schmiegelow},
  \citenamefont {Kaushal}, \citenamefont {Schulz}, \citenamefont {Mitchison},
  \citenamefont {Goold}, \citenamefont {Schmidt-Kaler},\ and\ \citenamefont
  {Poschinger}}]{vonlindenfels2019spin}%
  \BibitemOpen
  \bibfield  {author} {\bibinfo {author} {\bibfnamefont {D.}~\bibnamefont {von
  Lindenfels}}, \bibinfo {author} {\bibfnamefont {O.}~\bibnamefont {Gr\"ab}},
  \bibinfo {author} {\bibfnamefont {C.~T.}\ \bibnamefont {Schmiegelow}},
  \bibinfo {author} {\bibfnamefont {V.}~\bibnamefont {Kaushal}}, \bibinfo
  {author} {\bibfnamefont {J.}~\bibnamefont {Schulz}}, \bibinfo {author}
  {\bibfnamefont {M.~T.}\ \bibnamefont {Mitchison}}, \bibinfo {author}
  {\bibfnamefont {J.}~\bibnamefont {Goold}}, \bibinfo {author} {\bibfnamefont
  {F.}~\bibnamefont {Schmidt-Kaler}}, \ and\ \bibinfo {author} {\bibfnamefont
  {U.~G.}\ \bibnamefont {Poschinger}},\ }\href {\doibase
  10.1103/PhysRevLett.123.080602} {\bibfield  {journal} {\bibinfo  {journal}
  {Phys. Rev. Lett.}\ }\textbf {\bibinfo {volume} {123}},\ \bibinfo {pages}
  {080602} (\bibinfo {year} {2019})}\BibitemShut {NoStop}%
\bibitem [{\citenamefont {Peterson}\ \emph {et~al.}(2018)\citenamefont
  {Peterson}, \citenamefont {Batalh{\~a}o}, \citenamefont {Herrera},
  \citenamefont {Souza}, \citenamefont {Sarthour}, \citenamefont {Oliveira},\
  and\ \citenamefont {Serra}}]{peterson2018experimental}%
  \BibitemOpen
  \bibfield  {author} {\bibinfo {author} {\bibfnamefont {J.~P.~S.}\
  \bibnamefont {Peterson}}, \bibinfo {author} {\bibfnamefont {T.~B.}\
  \bibnamefont {Batalh{\~a}o}}, \bibinfo {author} {\bibfnamefont
  {M.}~\bibnamefont {Herrera}}, \bibinfo {author} {\bibfnamefont {A.~M.}\
  \bibnamefont {Souza}}, \bibinfo {author} {\bibfnamefont {R.~S.}\ \bibnamefont
  {Sarthour}}, \bibinfo {author} {\bibfnamefont {I.~S.}\ \bibnamefont
  {Oliveira}}, \ and\ \bibinfo {author} {\bibfnamefont {R.~M.}\ \bibnamefont
  {Serra}},\ }\href {\doibase 10.1103/PhysRevLett.123.240601} {\bibfield  {journal} {\bibinfo  {journal}
  {Phys. Rev. Lett.}\ }\textbf {\bibinfo {volume} {123}},\ \bibinfo {pages}
{240601} (\bibinfo {year} {2019})}\BibitemShut {NoStop}%
\bibitem [{\citenamefont {Klatzow}\ \emph {et~al.}(2019)\citenamefont
  {Klatzow}, \citenamefont {Becker}, \citenamefont {Ledingham}, \citenamefont
  {Weinzetl}, \citenamefont {Kaczmarek}, \citenamefont {Saunders},
  \citenamefont {Nunn}, \citenamefont {Walmsley}, \citenamefont {Uzdin},\ and\
  \citenamefont {Poem}}]{klatzow19experimental}%
  \BibitemOpen
  \bibfield  {author} {\bibinfo {author} {\bibfnamefont {J.}~\bibnamefont
  {Klatzow}}, \bibinfo {author} {\bibfnamefont {J.~N.}\ \bibnamefont {Becker}},
  \bibinfo {author} {\bibfnamefont {P.~M.}\ \bibnamefont {Ledingham}}, \bibinfo
  {author} {\bibfnamefont {C.}~\bibnamefont {Weinzetl}}, \bibinfo {author}
  {\bibfnamefont {K.~T.}\ \bibnamefont {Kaczmarek}}, \bibinfo {author}
  {\bibfnamefont {D.~J.}\ \bibnamefont {Saunders}}, \bibinfo {author}
  {\bibfnamefont {J.}~\bibnamefont {Nunn}}, \bibinfo {author} {\bibfnamefont
  {I.~A.}\ \bibnamefont {Walmsley}}, \bibinfo {author} {\bibfnamefont
  {R.}~\bibnamefont {Uzdin}}, \ and\ \bibinfo {author} {\bibfnamefont
  {E.}~\bibnamefont {Poem}},\ }\href {\doibase 10.1103/PhysRevLett.122.110601}
  {\bibfield  {journal} {\bibinfo  {journal} {Phys. Rev. Lett.}\ }\textbf
  {\bibinfo {volume} {122}},\ \bibinfo {pages} {110601} (\bibinfo {year}
  {2019})}\BibitemShut {NoStop}%
\bibitem [{\citenamefont {Scovil}\ and\ \citenamefont
  {Schulz-DuBois}(1959)}]{scovil1959three}%
  \BibitemOpen
  \bibfield  {author} {\bibinfo {author} {\bibfnamefont {H.~E.~D.}\
  \bibnamefont {Scovil}}\ and\ \bibinfo {author} {\bibfnamefont {E.~O.}\
  \bibnamefont {Schulz-DuBois}},\ }\href {\doibase 10.1103/PhysRevLett.2.262}
  {\bibfield  {journal} {\bibinfo  {journal} {Phys. Rev. Lett.}\ }\textbf
  {\bibinfo {volume} {2}},\ \bibinfo {pages} {262} (\bibinfo {year}
  {1959})}\BibitemShut {NoStop}%
\bibitem [{\citenamefont {Walls}\ and\ \citenamefont
  {Milburn}(1994)}]{wallsbook}%
  \BibitemOpen
  \bibfield  {author} {\bibinfo {author} {\bibfnamefont {D.~F.}\ \bibnamefont
  {Walls}}\ and\ \bibinfo {author} {\bibfnamefont {G.~J.}\ \bibnamefont
  {Milburn}},\ }\href@noop {} {\emph {\bibinfo {title} {Quantum Optics}}},\
  \bibinfo {edition} {1st}\ ed.\ (\bibinfo  {publisher} {Springer-Verlag},\
  \bibinfo {address} {Berlin},\ \bibinfo {year} {1994})\BibitemShut {NoStop}%
\bibitem [{\citenamefont {Niedenzu}\ \emph {et~al.}(2019)\citenamefont
  {Niedenzu}, \citenamefont {Huber},\ and\ \citenamefont
  {Boukobza}}]{niedenzu2019concepts}%
  \BibitemOpen
  \bibfield  {author} {\bibinfo {author} {\bibfnamefont {W.}~\bibnamefont
  {Niedenzu}}, \bibinfo {author} {\bibfnamefont {M.}~\bibnamefont {Huber}}, \
  and\ \bibinfo {author} {\bibfnamefont {E.}~\bibnamefont {Boukobza}},\ }\href
  {\doibase 10.22331/q-2019-10-14-195} {\bibfield  {journal} {\bibinfo
  {journal} {{Quantum}}\ }\textbf {\bibinfo {volume} {3}},\ \bibinfo {pages}
  {195} (\bibinfo {year} {2019})}\BibitemShut {NoStop}%
\bibitem [{\citenamefont {Pusz}\ and\ \citenamefont
  {Woronowicz}(1978)}]{pusz78}%
  \BibitemOpen
  \bibfield  {author} {\bibinfo {author} {\bibfnamefont {W.}~\bibnamefont
  {Pusz}}\ and\ \bibinfo {author} {\bibfnamefont {S.~L.}\ \bibnamefont
  {Woronowicz}},\ }\href {\doibase 10.1007/BF01614224} {\bibfield  {journal}
  {\bibinfo  {journal} {Commun. Math. Phys.}\ }\textbf {\bibinfo {volume}
  {58}},\ \bibinfo {pages} {273} (\bibinfo {year} {1978})}\BibitemShut
  {NoStop}%
\bibitem [{\citenamefont {Lenard}(1978)}]{lenard78}%
  \BibitemOpen
  \bibfield  {author} {\bibinfo {author} {\bibfnamefont {A.}~\bibnamefont
  {Lenard}},\ }\href {\doibase 10.1007/BF01011769} {\bibfield  {journal}
  {\bibinfo  {journal} {J. Stat. Phys.}\ }\textbf {\bibinfo {volume} {19}},\
  \bibinfo {pages} {575} (\bibinfo {year} {1978})}\BibitemShut {NoStop}%
\bibitem [{\citenamefont {Niedenzu}\ \emph {et~al.}(2018)\citenamefont
  {Niedenzu}, \citenamefont {Mukherjee}, \citenamefont {Ghosh}, \citenamefont
  {Kofman},\ and\ \citenamefont {Kurizki}}]{niedenzu18quantum}%
  \BibitemOpen
  \bibfield  {author} {\bibinfo {author} {\bibfnamefont {W.}~\bibnamefont
  {Niedenzu}}, \bibinfo {author} {\bibfnamefont {V.}~\bibnamefont {Mukherjee}},
  \bibinfo {author} {\bibfnamefont {A.}~\bibnamefont {Ghosh}}, \bibinfo
  {author} {\bibfnamefont {A.~G.}\ \bibnamefont {Kofman}}, \ and\ \bibinfo
  {author} {\bibfnamefont {G.}~\bibnamefont {Kurizki}},\ }\href {\doibase
  10.1038/s41467-017-01991-6} {\bibfield  {journal} {\bibinfo  {journal} {Nat.
  Commun.}\ }\textbf {\bibinfo {volume} {9}},\ \bibinfo {pages} {165} (\bibinfo
  {year} {2018})}\BibitemShut {NoStop}%
\bibitem [{\citenamefont {Binder}\ \emph {et~al.}(2015)\citenamefont {Binder},
  \citenamefont {Vinjanampathy}, \citenamefont {Modi},\ and\ \citenamefont
  {Goold}}]{binder15quantacell}%
  \BibitemOpen
  \bibfield  {author} {\bibinfo {author} {\bibfnamefont {F.~C.}\ \bibnamefont
  {Binder}}, \bibinfo {author} {\bibfnamefont {S.}~\bibnamefont
  {Vinjanampathy}}, \bibinfo {author} {\bibfnamefont {K.}~\bibnamefont {Modi}},
  \ and\ \bibinfo {author} {\bibfnamefont {J.}~\bibnamefont {Goold}},\ }\href
  {\doibase 10.1088/1367-2630/17/7/075015} {\bibfield  {journal} {\bibinfo
  {journal} {New Journal of Physics}\ }\textbf {\bibinfo {volume} {17}},\
  \bibinfo {pages} {075015} (\bibinfo {year} {2015})}\BibitemShut {NoStop}%
\bibitem [{\citenamefont {Campaioli}\ \emph {et~al.}(2017)\citenamefont
  {Campaioli}, \citenamefont {Pollock}, \citenamefont {Binder}, \citenamefont
  {C\'eleri}, \citenamefont {Goold}, \citenamefont {Vinjanampathy},\ and\
  \citenamefont {Modi}}]{campaioli17enhancing}%
  \BibitemOpen
  \bibfield  {author} {\bibinfo {author} {\bibfnamefont {F.}~\bibnamefont
  {Campaioli}}, \bibinfo {author} {\bibfnamefont {F.~A.}\ \bibnamefont
  {Pollock}}, \bibinfo {author} {\bibfnamefont {F.~C.}\ \bibnamefont {Binder}},
  \bibinfo {author} {\bibfnamefont {L.}~\bibnamefont {C\'eleri}}, \bibinfo
  {author} {\bibfnamefont {J.}~\bibnamefont {Goold}}, \bibinfo {author}
  {\bibfnamefont {S.}~\bibnamefont {Vinjanampathy}}, \ and\ \bibinfo {author}
  {\bibfnamefont {K.}~\bibnamefont {Modi}},\ }\href {\doibase
  10.1103/PhysRevLett.118.150601} {\bibfield  {journal} {\bibinfo  {journal}
  {Phys. Rev. Lett.}\ }\textbf {\bibinfo {volume} {118}},\ \bibinfo {pages}
  {150601} (\bibinfo {year} {2017})}\BibitemShut {NoStop}%
\bibitem [{\citenamefont {Gelbwaser-Klimovsky}\ and\ \citenamefont
  {Kurizki}(2015)}]{klimovsky15work}%
  \BibitemOpen
  \bibfield  {author} {\bibinfo {author} {\bibfnamefont {D.}~\bibnamefont
  {Gelbwaser-Klimovsky}}\ and\ \bibinfo {author} {\bibfnamefont
  {G.}~\bibnamefont {Kurizki}},\ }\href {\doibase 10.1038/srep07809} {\bibfield
   {journal} {\bibinfo  {journal} {Sci. Rep.}\ }\textbf {\bibinfo {volume}
  {5}},\ \bibinfo {pages} {7809} (\bibinfo {year} {2015})}\BibitemShut
  {NoStop}%
\bibitem [{\citenamefont {Gelbwaser-Klimovsky}\ and\ \citenamefont
  {Kurizki}(2014)}]{klimovsky14heat}%
  \BibitemOpen
  \bibfield  {author} {\bibinfo {author} {\bibfnamefont {D.}~\bibnamefont
  {Gelbwaser-Klimovsky}}\ and\ \bibinfo {author} {\bibfnamefont
  {G.}~\bibnamefont {Kurizki}},\ }\href {\doibase 10.1103/PhysRevE.90.022102}
  {\bibfield  {journal} {\bibinfo  {journal} {Phys. Rev. E}\ }\textbf {\bibinfo
  {volume} {90}},\ \bibinfo {pages} {022102} (\bibinfo {year}
  {2014})}\BibitemShut {NoStop}%
\bibitem [{\citenamefont {Kucsko}\ \emph
  {et~al.}(2013{\natexlab{b}})\citenamefont {Kucsko}, \citenamefont {Maurer},
  \citenamefont {Yao}, \citenamefont {Kubo}, \citenamefont {Noh}, \citenamefont
  {Lo}, \citenamefont {Park},\ and\ \citenamefont {Lukin}}]{kucsko13nanometre}%
  \BibitemOpen
  \bibfield  {author} {\bibinfo {author} {\bibfnamefont {G.}~\bibnamefont
  {Kucsko}}, \bibinfo {author} {\bibfnamefont {P.~C.}\ \bibnamefont {Maurer}},
  \bibinfo {author} {\bibfnamefont {N.~Y.}\ \bibnamefont {Yao}}, \bibinfo
  {author} {\bibfnamefont {M.}~\bibnamefont {Kubo}}, \bibinfo {author}
  {\bibfnamefont {H.~J.}\ \bibnamefont {Noh}}, \bibinfo {author} {\bibfnamefont
  {P.~K.}\ \bibnamefont {Lo}}, \bibinfo {author} {\bibfnamefont
  {H.}~\bibnamefont {Park}}, \ and\ \bibinfo {author} {\bibfnamefont {M.~D.}\
  \bibnamefont {Lukin}},\ }\href {\doibase 10.1038/nature12373} {\bibfield
  {journal} {\bibinfo  {journal} {Nature}\ }\textbf {\bibinfo {volume} {500}},\
  \bibinfo {pages} {54} (\bibinfo {year} {2013}{\natexlab{b}})}\BibitemShut
  {NoStop}%
\bibitem [{\citenamefont {Metcalf}\ and\ \citenamefont {van~der
  Straten}(2007)}]{metcalf07laser}%
  \BibitemOpen
  \bibfield  {author} {\bibinfo {author} {\bibfnamefont {H.~J.}\ \bibnamefont
  {Metcalf}}\ and\ \bibinfo {author} {\bibfnamefont {P.}~\bibnamefont {van~der
  Straten}},\ }in\ \href {\doibase 10.1002/9783527600441.oe005} {\emph
  {\bibinfo {booktitle} {The Optics Encyclopedia}}},\ \bibinfo {editor} {edited
  by\ \bibinfo {editor} {\bibfnamefont {T.~G.}\ \bibnamefont {Brown}}, \bibinfo
  {editor} {\bibfnamefont {K.}~\bibnamefont {Creath}}, \bibinfo {editor}
  {\bibfnamefont {H.}~\bibnamefont {Kogelnik}}, \bibinfo {editor}
  {\bibfnamefont {M.~A.}\ \bibnamefont {Kriss}}, \bibinfo {editor}
  {\bibfnamefont {J.}~\bibnamefont {Schmit}}, \ and\ \bibinfo {editor}
  {\bibfnamefont {M.~J.}\ \bibnamefont {Weber}}}\ (\bibinfo  {publisher}
  {WILEY-VCH Verlag},\ \bibinfo {year} {2007})\ pp.\ \bibinfo {pages}
  {975--1014}\BibitemShut {NoStop}%
\end{thebibliography}

%

\end{document}